\documentclass[12pt,draftclsnofoot,onecolumn]{IEEEtran}
\usepackage{amsfonts}
\usepackage[pdftex]{graphicx}
\usepackage{epsfig,latexsym}
\usepackage{float}
\usepackage{indentfirst}
\usepackage{color}

\usepackage{amssymb}
\usepackage{amsmath}
\usepackage{times}
\usepackage{subfigure}
\usepackage{psfrag}
\usepackage{cite}
\usepackage{booktabs}
\usepackage{multirow}
\usepackage{verbatim}
\usepackage{diagbox}

\newtheorem{theorem}{$\mathbf{Theorem}$}

\newtheorem{corollary}{$\mathbf{Corollary}$}

\newtheorem{remark}{$\mathbf{Remark}$}

\newcounter{tempEquationCounter} 
\newcounter{thisEquationNumber}

\begin{document}
	
	\title{Deep Learning for MIMO Channel Estimation: Interpretation, Performance, and Comparison}
	
	\author{Qiang~Hu,~Feifei~Gao,~\IEEEmembership{Fellow,~IEEE},~Hao Zhang,~Shi Jin,~\IEEEmembership{Senior~Member,~IEEE},~and~Geoffrey~Ye~Li,~\IEEEmembership{Fellow,~IEEE}
		\thanks{Q. Hu and H. Zhang are with the Department
			of Electrical Engineering, Tsinghua University, Beijing 100084, P. R. China (e-mail: huq16@mails.tsinghua.edu.cn; haozhang@mail.tsinghua.edu.cn).}
		\thanks{F. Gao is with the Institute for Artificial Intelligence, Tsinghua University (THUAI), 
			State Key Lab of Intelligent Technologies and Systems, Tsinghua University, 
			Beijing National Research Center for Information Science and Technology (BNRist), 
			Department of Automation, Tsinghua University,
			Beijing 100084, P. R. China (e-mail: feifeigao@ieee.org).}
		\thanks{S. Jin is with the National Mobile Communications Research Laboratory, Southeast University, Nanjing 210096, P. R. China (e-mail: jinshi@seu.edu.cn).}
		\thanks{G. Y. Li is with the School of Electrical and Computer Engineering, Georgia Institute of Technology, Atlanta, GA, USA (e-mail: liye@ece.gatech.edu).}

	}
	
	\maketitle
	\vspace{-10mm}
	\begin{abstract}
	Deep learning (DL) has emerged as an effective tool for channel estimation in wireless communication systems, especially under some imperfect environments. However, even with such unprecedented success, DL methods are often regarded as black boxes and are lack of explanations on their internal mechanisms, which severely limits further improvement and extension. In this paper, we present preliminary theoretical analysis on DL based channel estimation for multiple-antenna systems to understand and interpret its internal mechanism. Deep neural network (DNN) with rectified linear unit (ReLU) activation function is mathematically equivalent to a piecewise linear function. Hence, the corresponding DL estimator can achieve universal approximation to a large family of functions by making efficient use of piecewise linearity. We demonstrate that DL based channel estimation does not restrict to any specific signal model and approaches to the minimum mean-squared error (MMSE) estimation in various scenarios without requiring any prior knowledge of channel statistics. Therefore, DL based channel estimation outperforms or is at least comparable with traditional channel estimation, depending on the types of channels. Simulation results confirm the accuracy of the proposed interpretation and demonstrate the effectiveness of DL based channel estimation under both linear and nonlinear signal models.
	\end{abstract}
		
	
	\IEEEpeerreviewmaketitle
	
	\section{Introduction}
	\label{introduction}
	\IEEEPARstart{D}{eep} learning (DL) is making profound technological revolution to the concepts, patterns, methods and means of wireless communication systems~\cite{wang2017deep,gao2018comnet,yang2019hardware}. There have been many interesting results for the physical layer (PHY)~\cite{qin2019deep} or network layer of communications~\cite{liang2018}, including channel estimation~\cite{he2018deep,ye2018power}, channel state information (CSI) feedback compression~\cite{wang2018deep}, signal detection~\cite{o2017introduction}, and resource management~\cite{ye2019deep,liang2019deep}, etc. Among all DL applications to wireless communication systems, channel estimation is one of the most widely studied issues. The first attempt has been made in~\cite{ye2018power} to apply powerful DL methods to learn the characteristics of frequency selective wireless channels and combat the nonlinear distortion and interference for orthogonal frequency division multiplexing (OFDM) systems. In~\cite{huang2018deep}, a novel framework incorporates DL methods into massive multiple-input multiple-output (MIMO) systems to address direction-of-arrival (DoA) estimation and channel estimation problems. In~\cite{yang2019}, DL based channel estimation is extended to doubly selective channels and has numerically demonstrated better performance than the conventional estimators in many scenarios. In~\cite{soltani2019deep}, the channel matrix is regarded as an image and a DL based image super-resolution and denoising technique is employed to estimate the channel. Furthermore, a sparse complex-valued neural network structure is proposed in~\cite{yang2019MIMO} to tackle channel estimation in massive MIMO systems. Another branch of research attempts to establish a novel end-to-end deep neural networks (DNNs) architecture to replace all modules at the transmitter and at the receiver, respectively, instead of strengthening only certain modules~\cite{dorner2017deep,ye2018channel,ye2020deep}. 
	
	The above works have achieved great success across a variety of tasks over the recent years. Despite the rapid development of DL, the DNN embedded wireless communication system is generally considered as a black box for signal transmission/reception~\cite{qin2019deep}. Only numerical and experimental evaluations are available to demonstrate the powerful capability of DL in learning key functional components of wireless systems and there is nearly no analytical interpretation to confirm the advantages and disadvantages of DL methods when applied to communications. It is desired to understand why DL methods achieve astounding performance for a wide range of tasks for further performance improvement and extension to different environments. Moreover, the restrictions of DL methods to wireless communication systems are also very important for better understanding which scenarios are suitable for DL embedded communication systems.
	
	Another important issue is how well newly emerged data-driven DL methods are compared to the traditional expert-designed algorithms in the field of wireless communications~\cite{wen2018deep}. Impairments in PHY communications, such as noise, channel fading, interference, etc, have been thoroughly understood and addressed by well-established signal and coding theories from both practical and theoretical perspectives. It is yet unclear whether the black-box DL methods would be able to outperform the existing white-box approaches. In addition, the traditional ways of signal processing have been overturned by DL methods, in which satisfactory performance is still attainable in the absence of expert knowledge. Little research so far has dealt with how the DL methods learn from data and how the lack of expert knowledge affects the DL embedded communication systems. 
	
	There has been a rich history in addressing the complicated inner-workings of DL methods. The very first results have demonstrated the universal approximation of DNNs, that is, any continuous function defined on a compact set can be approximated at any precision using a DNN~\cite{cybenko1989approximation,hornik1989multilayer}. Recently, some effort has been made to analyze how the structural properties of DNNs, mainly depth and width, contribute to their powerful capability of modeling functions, i.e., expressive power of DNNs~\cite{Yoshua2011Shallow,bianchini2014complexity,telgarsky2016benefits}. These works converge to a similar conclusion that the expressive power of a DNN grows exponentially with its depth, which is compelling and provides some meaningful theoretical insights into superior performance of DNNs in practice. However, previous research seldom involves with the theoretical interpretation on the performance of DL embedded communication systems.

	Recently, more and more research has evidenced that DL methods are particularly suited to channel estimation. A comprehensive understanding of DL method's behavior on estimating channels is expected to provide guidance and inspiration for the further exploitation on DL based estimation theory. In this paper, we present an initial attempt on interpreting DL for channel estimation in multiple-antenna systems. Our contributions are listed as follows:
	\begin{itemize}
		\item Based on the property that a DNN with rectified linear unit (ReLU) activation function (ReLU DNN) is mathematically equivalent to a piecewise linear function, we derive a closed-form expression for the DL based channel estimation in linear systems.
		\item We analyze and compare the performance of the DL based channel estimation with the conventional methods, i.e., least-squared (LS) and linear minimum mean-squared error (LMMSE) estimators. We demonstrate that the DL estimator built on the DNNs with rectified linear unit (ReLU) activation function (ReLU DNNs) can well approximate the minimum mean-squared error (MMSE) estimator.
		\item We find and demonstrate that the DL estimator is highly sensitive to the quality of training data. Therefore, the performance of DL based channel estimation will significantly degrade if the statistics of training data mismatch the deployed environments.
	\end{itemize}

	The rest of this paper is organized as follows. The system model and the traditional channel estimation methods are introduced in Section~\ref{preliminary}. The DL based channel estimation is analyzed in Section~\ref{linear-sec}. Robustness of DL based channel estimation to mismatched training data is presented in Section~\ref{incomplete}. Simulation results are provided in Section~\ref{simulation} followed by the conclusions in Section~\ref{conclusion-sec}.
	
	\textit{Notations}: We use lowercase letters and capital letters in boldface to denote vectors and matrices, respectively. The positive integer set and real number set are denoted by $\mathbb{N}$ and $\mathbb{R}$, respectively. $\mathbf{I}_{M}$ denotes the $M\times M$ identity matrix. Notation $(\cdot)^{T}$ represent the transpose of a matrix or a vector, respectively. $\mathbb{E}\{\cdot\}$ denotes the expectation and $\mathrm{tr}\{\cdot\}$ denotes the trace of a matrix. The cardinality of a set is denoted by $|\cdot|$. Notation $\|\cdot\|_{2}$ represents the $2$-norm of a vector. Notation $\lceil\cdot\rceil$ represents the ceiling of a real number.  
	
	\section{System Model and Traditional Channel Estimation}\label{preliminary}
	In this section, we first introduce the multiple-antenna communication system for channel estimation and then present the traditional channel estimation methods. 
	\subsection{System Model}
	Consider a multiple-antenna communication system with $d$ antennas at the base station (BS) and a single antenna at the user side. Assume the uplink channel is with block fading, that is, channel parameters are fixed within a block but vary from one to another. The traditional way of estimating channels at the BS is to use uplink pilot. Let $\tau$ be the transmitted pilot symbol with $|\tau|^{2}=1$. The received symbol at the BS can be represented by the following $d\times 1$ vector 
	\begin{align}\label{r_train}
	\mathbf{x}=\tau\mathbf{h}+\mathbf{n},
	\end{align}
	where $\mathbf{h}$ denotes the $d\times 1$ random channel vector between the user and the BS and $\mathbf{n}$ is the $d\times 1$ white Gaussian noise vector with zero-mean and element-wise variance $\sigma_{n}^{2}$. We assume that the channel vector $\mathbf{h}$ is with zero mean and covariance matrix $\mathbf{R}=\mathbb{E}\{\mathbf{h}\mathbf{h}^{T}\}$.\footnote{In general, the complex valued signals would decompose into real values before inputting to ReLU DNNs. For convenience, we assume that both of input signals and channels are real values.}
	
	\subsection{Traditional Channel Estimation}
	
	The goal of channel estimation is to extract channel vector $\mathbf{h}$ from received signal vector $\mathbf{x}$ as accurately as possible. The traditional estimation methods are based on the signal model in~\eqref{r_train}. 
	
	\subsubsection{LS Channel Estimator}
	From~\eqref{r_train}, the LS estimate of $\mathbf{h}$ can be expressed as~\cite{kay1993fundamentals} 
	\begin{align}
	\mathbf{h}_{\mathrm{LS}}=\frac{1}{\tau}\mathbf{x}=\mathbf{h}+\frac{1}{\tau}\mathbf{n},
	\end{align}   
	and the corresponding MSE is
	\begin{align}\label{ls3}
	J_{\mathrm{LS}}&=\mathbb{E}\{\|\mathbf{h}-\mathbf{h}_{\mathrm{LS}}\|^{2}_{2}\}=\frac{d}{1/\sigma_{n}^{2}}.
	\end{align}
	
	As shown in~\eqref{ls3}, the performance of the LS estimator is inversely proportional to the signal-to-noise ratio (SNR) defined as $1/\sigma_{n}^{2}$.

	\subsubsection{LMMSE Channel Estimator}
	
	The LMMSE estimator exploits the signal model in~\eqref{r_train} and channel statistics to obtain the estimation, which can be expressed as~\cite{kay1993fundamentals}
	\begin{align}\label{mmse1}
	\mathbf{h}_{\mathrm{LMMSE}}&=\mathbf{R}\left(\mathbf{R}+\sigma_{n}^{2}\mathbf{I}_{d}\right)^{-1}\mathbf{x}.
	\end{align}
	Then, the MSE of the LMMSE estimator is computed as
	\begin{align}
	J_{\mathrm{LMMSE}}=\mathrm{tr}\bigg\{\mathbf{R}\left(\mathbf{I}_{d}+\frac{1}{\sigma_{n}^{2}}\mathbf{R}\right)^{-1}\bigg\}\leq J_{\mathrm{LS}}.
	\end{align}
	
	\subsubsection{MMSE Channel Estimator}
	
	 The MMSE estimator can be expressed as~\cite{kay1993fundamentals}
	 \begin{align}\label{mmse}
	 \mathbf{h}_{\mathrm{MMSE}}=\mathbb{E}\{\mathbf{h}|\mathbf{x}\}
	 \end{align}
	and is optimal under the criterion of minimizing the MSE. In general, the MMSE estimator is different from the LMMSE estimator. Specifically, it holds that $\mathbf{h}_{\mathrm{MMSE}}=\mathbf{h}_{\mathrm{LMMSE}}$ if $\mathbf{x}$ and $\mathbf{h}$ are joint Gaussian distributed for linear models, such as in~\eqref{r_train}. Therefore, the LS and LMMSE estimators can well address the channel estimation for linear models as in~\eqref{r_train}. However, both LS and LMMSE estimators are linear and their estimation performance degrades significantly for nonlinear models. As a result, the MSE of $\mathbf{h}_{\mathrm{MMSE}}$ is usually no larger than that of $\mathbf{h}_{\mathrm{LMMSE}}$. 
	
	The LMMSE and MMSE channel estimation leads to a more accurate estimate by utilizing the statistics of channels, and therefore the performance is sensitive to the imperfection of channel statistics. On the contrary, the LS estimator is easy to implement due to no prior requirement on channel statistics, but such simplicity is at cost of relatively low accuracy. 
	
	Recently, the DL estimator has emerged as a promising alternative to address channel estimation in wireless communication systems. The excellent generalization ability and powerful learning capacity of the DL estimator make it a powerful tool for channel estimation for imperfect and interference corrupted systems.

	\section{Analysis on DL based Channel Estimation}\label{linear-sec}
Though DL based channel estimation has shown excellent performance in various communication systems, it has seldom been analyzed from a theoretical perspective. In this section, we provide a preliminary theoretical analysis on the performance of the DL based channel estimation via statistical learning theory will be provided. Specifically, we demonstrate the DL estimator asymptotically approaches to the MMSE estimator as the number of training sample increases. 
	\subsection{Basic Setting of DL Channel Estimator}
	Consider a DL estimator $\mathcal{D}$ with a  fully-connected DNN with ReLU activation function. The input and output of $\mathcal{D}$ are denoted by $\mathbf{x}\in\mathbb{R}^{d}$ and $\mathbf{h}\in\mathbb{R}^{d}$, respectively. In the subsequent discussion, we will denote 
	\begin{align}
	\mathcal{Z}=\{(\mathbf{x}_{m},\mathbf{h}_{m})\,|\,\mathbf{x}_{m},\mathbf{h}_{m}\in\mathbb{R}^{d},\,m=1,\ldots,|\mathcal{Z}|\}
	\end{align}	
	as input-output sample set. 
	
	The underlying DNN of  $\mathcal{D}$  consists of  the ReLU activation function, $\varphi(x)=\max\{0,x\}$, a number of layers $l\in\mathbb{N}$, and the neuron assignment $\mathbf{d}=(d_{0},d_{1},\ldots,d_{l},d_{l+1})\in\mathbb{N}^{l+2}$ with $d_{0}=d_{l+1}=d$. The number of hidden layers $l$ is referred to as the depth of $\mathcal{D}$. The width and size of $\mathcal{D}$ are denoted by $\max\{d_{1},\ldots,d_{l}\}$ and $\sum_{i=1}^{l}d_{l}$, respectively. 
	
	Let
	 \begin{align}
	 \Theta=\big\{\boldsymbol{\theta}=\left\{(\mathbf{W}_{i},\mathbf{b}_{i})\right\}_{i=0}^{l}\in\{\mathbb{R}^{d_{i+1}\times (d_{i}+1)}\}_{i=0}^{l}\big\}
	 \end{align} be
	the set of all parameters of $\mathcal{D}$, where $\mathbf{W}_{i}\in\mathbb{R}^{d_{i+1}\times d_{i}}$ and $\mathbf{b}_{i}\in\mathbb{R}^{d_{i+1}}$ are the weight matrix and the bias vector of the $i$-th layer. For a fixed network structure $\mathbf{d}$, the underlying function that represented by $\mathcal{D}$  can be expressed as
	\begin{align}\label{network1}
	\mathbf{f}_{\boldsymbol{\theta}}(\mathbf{x})=\mathcal{A}_{l}\circ\varphi_{d_{l}}\circ\mathcal{A}_{l-1}\circ\varphi_{d_{l-1}}\circ\cdots\circ\varphi_{d_{1}}\circ\mathcal{A}_{0}(\mathbf{x}),
	\end{align} 
	where $\mathcal{A}_{i}:\mathbb{R}^{d_{i}}\rightarrow\mathbb{R}^{d_{i+1}}$ is the affine transformation corresponding to weight $\mathbf{W}_{i}$ and bias $\mathbf{b}_{i}$, $\varphi_{d_{i}}:\mathbb{R}^{d_{i}}\rightarrow\mathbb{R}^{d_{i}}$ is the entry-wise ReLU activation function, and $\circ$ denotes the function composition. 
	The goal of DL based channel estimation is to adjust $\boldsymbol{\theta}$ in order to approximate the MMSE estimation functional relationship between $\mathbf{x}$ and $\mathbf{h}$ for given training sample set $\mathcal{Z}$ and network architecture $\mathbf{d}$. 
	
	\subsection{Internal Mechanism of DL Channel Estimator}

	Since ReLU function is piecewise linear, the neurons in $\mathcal{D}$ consist of only two states: with zero output or replicating input. When $\boldsymbol{\theta}$ is fixed, all the possible activation patterns of neurons in $\mathcal{D}$ can be represented by a set $\mathcal{K}\subseteq\{0,1\}^{\bar{d}}$, where $\bar{d}=\sum_{i=1}^{l}d_{i}$ is a total number of neurons in $\mathcal{D}$ and each element in $\mathcal{K}$ is a $\bar{d}$-dimensional vector with its entries either $0$ or $1$. It is obvious that $|\mathcal{K}|$ is upper bounded by $2^{\bar{d}}$, i.e, $|\mathcal{K}|\leq 2^{\bar{d}}$.	Similar to~\cite{montufar2014number}, the input space of a ReLU DNN is partitioned into different linear regions according to the corresponding activation patterns so that the ReLU DNN turns into a linear mapping in each region.  Denote $\mathcal{X}$ as the input space and $\mathcal{X}_{k}$ as the input region within $\mathcal{X}$ corresponding to the $k$-th activation pattern. It is obvious that\footnote{These linear regions are not necessarily disjoint.}
	\begin{align}\label{partition}
	&\mathcal{X}_{k}\subseteq\mathcal{X},\ k=1,\ldots,K=|\mathcal{K}|,\nonumber\\
	&\mathcal{X}=\cup_{k=1}^{K}\mathcal{X}_{k}.
	\end{align}

	Let $\tilde{\mathbf{x}}_{i}=[x_{i,1},\ldots,x_{i,d_{i}}]^{T}$ be the output of the $i$-th layer with $\tilde{\mathbf{x}}_{0}=\mathbf{x}$. For any input $\mathbf{x}\in\mathcal{X}_{k}$, $\mathcal{A}_{i}(\tilde{\mathbf{x}}_{i})$ in~\eqref{network1} is computed as
	\begin{align}\label{hyperplane1}
	\mathcal{A}_{i}(\tilde{\mathbf{x}}_{i})=\left\{\begin{array}{lcl}\mathbf{W}_{0}\mathbf{x}+\mathbf{b}_{0}, & & i=0,\\
	\tilde{\mathbf{W}}_{i}\mathcal{A}_{i-1}(\tilde{\mathbf{x}}_{i-1})+\mathbf{b}_{i},& &i\geq 1,\end{array}\right.
	\end{align}
	where $\tilde{\mathbf{W}}_{i}=\mathbf{W}_{i}\mathbf{\Lambda}_{i}$ and $\mathbf{\Lambda}_{i}$ is a $\mathbb{R}^{d_{i}\times d_{i}}$ diagonal matrix with the diagonal elements either $0$ or $1$. Note that $\mathbf{\Lambda}_{0}=\mathbf{I}_{d}$ and  $\tilde{\mathbf{W}}_{0}=\mathbf{W}_{0}\mathbf{\Lambda}_{0}$. Moreover, the diagonal elements of  $\mathbf{\Lambda}_{i}$ correspond to the activation pattern of neurons at the $i$-th layer with their values either $0$ or $1$. Since all inputs $\mathbf{x}\in\mathcal{X}_{k}$ have the same activation pattern, the set $\{\mathbf{\Lambda}_{i}\}_{i=0}^{l}$ is fixed. By recursively expanding $\mathcal{A}_{i}(\tilde{\mathbf{x}}_{i})$ layer by layer, we can further express $\mathcal{A}_{i}(\tilde{\mathbf{x}}_{i})$ as
	\begin{align}\label{hyperplane}
	\mathcal{A}_{i}(\tilde{\mathbf{x}}_{i})&=\prod_{j=0}^{i}\tilde{\mathbf{W}}_{j}\mathbf{x}+\sum_{j=0}^{i-1}\bigg(\prod_{p=0}^{j}\tilde{\mathbf{W}}_{j+1-p}\bigg)\mathbf{b}_{i-1-j}+\mathbf{b}_{i}\nonumber\\
	&=\hat{\mathbf{W}}_{i}\mathbf{x}+\hat{\mathbf{b}}_{i},
	\end{align}
	where $\hat{\mathbf{W}}_{i}=\prod_{j=0}^{i}\tilde{\mathbf{W}}_{j}$ is the equivalent weight matrix with respect to (w.r.t) the input $\mathbf{x}$ and $\hat{\mathbf{b}}_{i}=\sum_{j=0}^{i-1}\left(\prod_{p=0}^{j}\tilde{\mathbf{W}}_{j+1-p}\right)\mathbf{b}_{i-1-j}+\mathbf{b}_{i}$ is the sum of the remaining terms. Therefore, $\mathbf{f}_{\boldsymbol{\theta}}\left(\mathbf{x}\right)$ turns into an affine function	for $\mathbf{x}\in\mathcal{X}_{k}$ and can be expressed as
	\begin{align}\label{network} \mathbf{f}_{\boldsymbol{\theta}}\left(\mathbf{x}\right)=\mathbf{f}_{\mathcal{X}_{k}}\left(\mathbf{x}\right)=\mathbf{W}_{\mathcal{X}_{k}}\mathbf{x}+\mathbf{b}_{\mathcal{X}_{k}},
	\end{align}
	where $\mathbf{W}_{\mathcal{X}_{k}}=\hat{\mathbf{W}}_{l}$ and $\mathbf{b}_{\mathcal{X}_{k}}=\hat{\mathbf{b}}_{l}$. 
\begin{remark}\label{remark1}
	\begin{itemize}
		\item If $\boldsymbol{\theta}$ is fixed, the inputs belonging to the same region have the same form of $\mathbf{f}_{\mathcal{X}_{k}}\left(\mathbf{x}\right)$ and correspond to the same activation pattern. From~\eqref{network}, $\mathbf{f}_{\mathcal{X}_{k}}\left(\mathbf{x}\right)$ is an affine function for $\mathbf{x}\in\mathcal{X}_{k}$, and therefore $\mathbf{f}_{\boldsymbol{\theta}}\left(\mathbf{x}\right)$ is a $\mathbb{R}^{d}\rightarrow\mathbb{R}^{d}$ piecewise linear function over $\mathcal{X}$.
		\item The DL estimator has the simplicity and stability of a linear estimator, but also enables remarkable flexibility through the piecewise linear property. 
		\item The DL estimator can model a large family of nonlinear functions by dynamically adjusting the partitioned regions,and is more general and flexible compared to the LS and LMMSE estimators. The piecewise linear property of $\mathbf{f}_{\boldsymbol{\theta}}(\mathbf{x})$ is a critical step to interpret DL based channel estimation and will be used in the later analysis.
	\end{itemize}
\end{remark}
	
	\subsection{Performance Assessment of DL Channel Estimator}
	Different from the LS and LMMSE estimators, it is difficult to derive an explicit analytical form of the DL estimate as well as the corresponding MSE. Hence, the performance assessment of the DL estimator and the comparison to the LS and LMMSE estimators are not straightforward. Nevertheless, the DL estimator can approximate to a large family of functions due to its piecewise linear property. We can leverage the universal approximation of the DL estimator to assess its estimation performance and prove its convergence to the MMSE estimator. 
	
	Let  $
	\mathbf{f}(\mathbf{x}):\mathbb{R}^{d}\rightarrow\mathbb{R}^{d}$ denote a Lebesgue measurable function for estimating $\mathbf{h}$ and $\ell_{2}$ be the space of  all measurable functions $\mathbf{f}(\mathbf{x})$ with finite $\ell_{2}$-norm defined as
	\begin{align}
	 \|\mathbf{f}(\mathbf{x})\|_{\ell_{2}}=\bigg[\sum_{i=1}^{d}\mathbb{E}\left\{\|f_{i}(\mathbf{x})\|_{2}^{2}\right\}\bigg]^{1/2}<+\infty,
	\end{align} where $f_{i}(\mathbf{x})$ is the $i$-th entry of $\mathbf{f}(\mathbf{x})$. The expected loss or the MSE of $\mathbf{f}(\mathbf{x})$ is defined as
	\begin{align}\label{optimal1}
	J(\mathbf{f})&=\mathbb{E}\{\|\mathbf{f}(\mathbf{x})-\mathbf{h}\|^{2}_{2}\}.
	\end{align} 
	
	Denote
	\begin{align}
	\mathbf{f}_{o}(\mathbf{x})=\mathbb{E}\{\mathbf{h}|\mathbf{x}\},
	\end{align}
	and we assume that $\mathbf{f}_{o}(\mathbf{x})\in \ell_{2}$ throughout this paper. From the orthogonal principle, we have
	\begin{align}\label{optimal}
	J(\mathbf{f})&=\mathbb{E}\big\{\|\mathbf{f}(\mathbf{x})-\mathbf{f}_{o}(\mathbf{x})+\mathbf{f}_{o}(\mathbf{x})-\mathbf{h}\|_{2}^{2}\big\}\nonumber\\
	&=\mathbb{E}\big\{\|\mathbf{f}(\mathbf{x})-\mathbf{f}_{o}(\mathbf{x})\|_{2}^{2}\big\}+\mathbb{E}\big\{\|\mathbf{f}_{o}(\mathbf{x})-\mathbf{h}\|_{2}^{2}\big\}+2\mathbb{E}\big\{(\mathbf{f}(\mathbf{x})-\mathbf{f}_{o}(\mathbf{x}))^{T}(\mathbf{f}_{o}(\mathbf{x})-\mathbf{h})\big\}\nonumber\\
	&=\mathbb{E}\big\{\|\mathbf{f}(\mathbf{x})-\mathbf{f}_{o}(\mathbf{x})\|_{2}^{2}\big\}+J(\mathbf{f}_{o}).
	\end{align}
	The first term in the right-hand side (RHS) of~\eqref{optimal} is the expectation of the squared $2$-norm distance from the use of $\mathbf{f}(\mathbf{x})$ to model $\mathbf{f}_{o}(\mathbf{x})$ and is non-negative. In this sense, the second term $J(\mathbf{f}_{o})$ in the RHS of~\eqref{optimal} provides a lower bound on the expected loss $J(\mathbf{f})$, which is independent of $\mathbf{f}$ and is determined by the joint distribution of $\mathbf{x}$ and $\mathbf{h}$. As a result,  
	\begin{align}
	\mathbf{f}_{o}(\mathbf{x})=\arg \min_{\mathbf{f}(\mathbf{x})\in \ell_{2}} J(\mathbf{f}),
	\end{align}
	that is, $\mathbf{f}_{o}(\mathbf{x})$ is the MMSE estimation w.r.t. $\mathbf{h}$, as we have indicated in Section~\ref{preliminary}, and $J(\mathbf{f}_{o})$ is the corresponding MSE of the MMSE estimator. 
	
	Specifically, if  both $\mathbf{x}$ and $\mathbf{h}$ are Gaussian distributed for linear models as in~\eqref{r_train},  we have $\mathbf{h}_{\mathrm{MMSE}}=\mathbf{h}_{\mathrm{LMMSE}}$ and $J(\mathbf{f}_{o})$ is simply equivalent to $J_{\mathrm{LMMSE}}$ . 
	
	Given a ReLU DNN with parameter $\boldsymbol{\theta}$, the input-output relation can be expressed as a function $\mathbf{f}_{\boldsymbol{\theta}}(\mathbf{x})$. Let
	\begin{align}\label{expected}
	J(\mathbf{f}_{\boldsymbol{\theta}})=\mathbb{E}\big\{\|\mathbf{f}_{\boldsymbol{\theta}}(\mathbf{x})-\mathbf{h}\|_{2}^{2}\big\}
	\end{align}
	be the expected loss and 
	\begin{align}\label{empirical}
	J_{\mathcal{Z}}(\mathbf{f}_{\boldsymbol{\theta}})=\frac{1}{|\mathcal{Z}|}\sum_{(\mathbf{x}_{m},\mathbf{h}_{m})\in\mathcal{Z}}\|\mathbf{f}_{\boldsymbol{\theta}}(\mathbf{x}_{m})-\mathbf{h}_{m}\|_{2}^{2}
	\end{align}
	be the empirical loss w.r.t. $\mathcal{Z}$, respectively. 
	
	Denote
	\begin{align}\label{mini1}
	\boldsymbol{\theta}_{o}=\arg\min_{\boldsymbol{\theta}\in\Theta}J(\mathbf{f}_{\boldsymbol{\theta}}),\ J(\mathbf{f}_{\boldsymbol{\theta}_{o}})=\min_{\boldsymbol{\theta}\in\Theta}J(\mathbf{f}_{\boldsymbol{\theta}}).
	\end{align}
	That is, $\mathbf{f}_{\boldsymbol{\theta}_{o}}(\mathbf{x})$ is the optimal DL estimator for a DNN with the given structure, and $\boldsymbol{\theta}_{o}$ is the corresponding parameter of the DNN. It is obvious that $J(\mathbf{f}_{\boldsymbol{\theta}_{o}})$ is the minimum MSE over all the DL estimators. 
	
	Similarly, denote
	\begin{align}\label{min2}
	\boldsymbol{\theta}_{\mathcal{Z}}=\arg\min_{\boldsymbol{\theta}\in\Theta}J_{\mathcal{Z}}(\mathbf{f}_{\boldsymbol{\theta}}),\ J_{\mathcal{Z}}(\mathbf{f}_{\boldsymbol{\theta}_{\mathcal{Z}}})=\min_{\boldsymbol{\theta}\in\Theta}J_{\mathcal{Z}}(\mathbf{f}_{\boldsymbol{\theta}}).
	\end{align}
	In the above, $\mathbf{f}_{\boldsymbol{\theta}_{\mathcal{Z}}}(\mathbf{x})$ is the optimal DL estimator with a DNN trained by the dataset $\mathcal{Z}$ using the empirical loss as in~\eqref{empirical}. Therefore,
	\begin{align}\label{dlesti}
	\mathbf{h}_{\mathrm{DL}}=\mathbf{f}_{\boldsymbol{\theta}_{\mathcal{Z}}}(\mathbf{x})
	\end{align} 
	can be regarded as the best DL estimator that can be obtained in practice. The corresponding expected loss or MSE of the DL estimator will be $J(\mathbf{f}_{\boldsymbol{\theta}_{\mathcal{Z}}})$, which is obviously no less than $J(\mathbf{f}_{\boldsymbol{\theta}_{o}})$, that is, $J(\mathbf{f}_{\boldsymbol{\theta}_{\mathcal{Z}}})-J(\mathbf{f}_{\boldsymbol{\theta}_{o}})\geq 0$. 
	
	Let us then analyze the performance of the DL estimator through quantifying the expected loss of $\mathbf{f}_{\boldsymbol{\theta}_{\mathcal{Z}}}(\mathbf{x})$, i.e., $J(\mathbf{f}_{\boldsymbol{\theta}_{\mathcal{Z}}})$ in the linear and nonlinear systems.

	\subsubsection{Linear Systems}
	
	For the linear model in~\eqref{r_train} with $\mathbf{h}$ being Gaussian, $\mathbf{h}_{\mathrm{MMSE}}$ is simply equivalent to $\mathbf{h}_{\mathrm{LMMSE}}$, i.e., $\mathbf{f}_{o}(\mathbf{x})=\mathbf{h}_{\mathrm{LMMSE}}$. Let $\mathcal{A}:\mathbb{R}^{d}\rightarrow\mathbb{R}^{d}$ denote the affine transformation with weight $\mathbf{W}\in\mathbb{R}^{d\times d}$ and bias $\mathbf{b}\in\mathbb{R}^{d}$. Based on the property that the ReLU DNN can represent any affine transformation $\mathcal{A}$, the following theorem provides the explicit form of $\mathbf{f}_{\boldsymbol{\theta}_{\mathcal{Z}}}(\mathbf{x})$ in the linear systems and demonstrates that the expected loss of the DL estimator, i.e., $J(\mathbf{f}_{\boldsymbol{\theta}_{\mathcal{Z}}})$, asymptotically approaches to $J_{\mathrm{LMMSE}}$. 
	
	\begin{theorem}\label{theorem1-1}
		If $\mathbf{f}_{o}(\mathbf{x})=\mathbf{h}_{\mathrm{LMMSE}}$, then there exits an optimized DL estimator $\mathbf{f}_{\boldsymbol{\theta}_{\mathcal{Z}}}(\mathbf{x})$ built on the $2$-layer ReLU DNN of size $2d$ such that
		\begin{align}\label{theo1-1}
		J(\mathbf{f}_{\boldsymbol{\theta}_{\mathcal{Z}}})\stackrel{\mathbf{P}}{\longrightarrow}J_{\mathrm{LMMSE}},
		\end{align}
		that is 
		\begin{align}
	\lim_{|\mathcal{Z}|\rightarrow+\infty}\mathbf{P}\big(\|
	\mathbf{f}_{\boldsymbol{\theta}_{\mathcal{Z}}}(\mathbf{x})-\mathbf{h}_{\mathrm{LMMSE}}
	\|_{2}>\varepsilon\big)=0
	\end{align}
	for any $\varepsilon>0$, where $\mathbf{P}$ denotes  the distribution of the training sample in $\mathcal{Z}$. The explicit form of  $\mathbf{f}_{\boldsymbol{\theta}_{\mathcal{Z}}}(\mathbf{x})$ with sufficiently large $|\mathcal{Z}|$ is given by
	\begin{align}\label{theo2-1}
	\mathbf{f}_{\boldsymbol{\theta}_{\mathcal{Z}}}(\mathbf{x})&=\bigg(\sum_{m=1}^{|\mathcal{Z}|}\bar{\mathbf{h}}_{m}\bar{\mathbf{x}}_{m}^{T}\bigg)\bigg(\sum_{m=1}^{|\mathcal{Z}|}\bar{\mathbf{x}}_{m}\bar{\mathbf{x}}_{m}^{T}\bigg)^{-1}(\mathbf{x}-\mathbf{x}_{\mathcal{Z}})+\mathbf{h}_{\mathcal{Z}},
	\end{align}
	where $\mathbf{x}_{\mathcal{Z}}=\frac{1}{|\mathcal{Z}|}\sum_{m=1}^{|\mathcal{Z}|}\mathbf{x}_{m}$, $\mathbf{h}_{\mathcal{Z}}=\frac{1}{|\mathcal{Z}|}\sum_{m=1}^{|\mathcal{Z}|}\mathbf{h}_{m}$, $\bar{\mathbf{x}}_{m}=\mathbf{x}_{m}-\mathbf{x}_{\mathcal{Z}}$ and $\bar{\mathbf{h}}_{m}=\mathbf{h}_{m}-\mathbf{h}_{\mathcal{Z}}$.
	\end{theorem}
	\begin{IEEEproof}
		Due to the fact that
		\begin{align}\label{affine} \mathcal{A}=(\mathbf{I}_{d}\circ\varphi_{d}\circ\mathcal{A})+(-\mathbf{I}_{d}\circ\varphi_{d}\circ(-\mathcal{A})),
		\end{align}
		the RHS of~\eqref{affine} is equivalent to a $2$-layer ReLU DNN of size $2d$~\cite[Lemma D.4.]{arora2018understanding}, and therefore $\mathcal{A}$ is representable by the ReLU DNN. Let $\mathcal{A}(\mathbf{x})=\mathbf{h}_{\mathrm{LMMSE}}$, and there exits a DL estimator to represent $\mathbf{h}_{\mathrm{LMMSE}}$. From~\eqref{min2}, $\mathbf{f}_{\boldsymbol{\theta}_{\mathcal{Z}}}(\mathbf{x})$ has the lowest empirical loss over all the DL estimators, and we have
		\begin{align}\label{app1-2}
		 J_{\mathcal{Z}}(\mathbf{f}_{\boldsymbol{\theta}_{\mathcal{Z}}})\leq J_{\mathcal{Z}}(\mathcal{A}).
		\end{align}
		
		According to the law of large numbers~\cite{mendenhall2012introduction}, there is
		\begin{align}\label{affine1}
		J_{\mathcal{Z}}(\mathbf{f}_{\boldsymbol{\theta}_{\mathcal{Z}}})\stackrel{\mathbf{P}}{\longrightarrow}J(\mathbf{f}_{\boldsymbol{\theta}_{\mathcal{Z}}})\ \mathrm{and}\ J_{\mathcal{Z}}(\mathcal{A})\stackrel{\mathbf{P}}{\longrightarrow}J_{\mathrm{LMMSE}}.
		\end{align}
		Then, we can rewrite~\eqref{app1-2} as
		\begin{align}\label{app1-3}
		 J(\mathbf{f}_{\boldsymbol{\theta}_{\mathcal{Z}}})\leq J_{\mathrm{LMMSE}}
		\end{align}
		when $|\mathcal{Z}|$ is sufficiently large.
		Since $J_{\mathrm{LMMSE}}$ has the lowest MSE in the linear systems, and therefore
		\begin{align}\label{app1-4}
		J_{\mathrm{LMMSE}}\leq J(\mathbf{f}_{\boldsymbol{\theta}_{\mathcal{Z}}}).
		\end{align}
		Together with~\eqref{app1-3}, we have
		\begin{align}\label{app1-5}
		 J(\mathbf{f}_{\boldsymbol{\theta}_{\mathcal{Z}}})=J_{\mathrm{LMMSE}}\ \mathrm{and}\ J(\mathbf{f}_{\boldsymbol{\theta}_{\mathcal{Z}}})\stackrel{\mathbf{P}}{\longrightarrow}J_{\mathrm{LMMSE}}.
		 \end{align} 
		 
		 From~\eqref{network}, $\mathbf{f}_{\boldsymbol{\theta}_{\mathcal{Z}}}(\mathbf{x})$ is equivalent to the piecewise linear function. According to~\eqref{app1-5}, $\mathbf{f}_{\boldsymbol{\theta}_{\mathcal{Z}}}(\mathbf{x})$ reduces to an affine function when $|\mathcal{Z}|$ is sufficiently large. The number of partitioned regions of $\mathbf{f}_{\boldsymbol{\theta}_{\mathcal{Z}}}(\mathbf{x})$ is also equal to $1$, i.e., $K=1$. Then, minimizing $J_{\mathcal{Z}}(\mathbf{f}_{\boldsymbol{\theta}_{\mathcal{Z}}})$ with sufficiently large $|\mathcal{Z}|$ yields the affine function $\mathcal{A}(\mathbf{x})$ with \begin{align}\label{af-weight}
		 \mathbf{W}=\Big(\sum_{m=1}^{|\mathcal{Z}|}\bar{\mathbf{h}}_{m}
		 \bar{\mathbf{x}}^{T}_{m}\Big)\Big(\sum_{m=1}^{|\mathcal{Z}|}\bar{\mathbf{x}}_{m}
		 \bar{\mathbf{x}}^{T}_{m}\Big)^{-1}
		 \end{align}
		 and 
		 \begin{align}\label{af-bias}
		 \mathbf{b}= \frac{1}{|\mathcal{Z}|}\sum_{m=1}^{|\mathcal{Z}|}(\mathbf{h}_{m}-\mathbf{W}\mathbf{x}_{m}),
		 \end{align}
		 i.e.,
		\begin{align}
		\mathbf{f}_{\boldsymbol{\theta}_{\mathcal{Z}}}(\mathbf{x})&=\bigg(\sum_{m=1}^{|\mathcal{Z}|}\bar{\mathbf{h}}_{m}\bar{\mathbf{x}}_{m}^{T}\bigg)\bigg(\sum_{m=1}^{|\mathcal{Z}|}\bar{\mathbf{x}}_{m}\bar{\mathbf{x}}_{m}^{T}\bigg)^{-1}(\mathbf{x}-\mathbf{x}_{\mathcal{Z}})+\mathbf{h}_{\mathcal{Z}},
		\end{align}
		where $\mathbf{x}_{\mathcal{Z}}=\frac{1}{|\mathcal{Z}|}\sum_{m=1}^{|\mathcal{Z}|}\mathbf{x}_{m}$, $\mathbf{h}_{\mathcal{Z}}=\frac{1}{|\mathcal{Z}|}\sum_{m=1}^{|\mathcal{Z}|}\mathbf{h}_{m}$, $\bar{\mathbf{x}}_{m}=\mathbf{x}_{m}-\mathbf{x}_{\mathcal{Z}}$ and $\bar{\mathbf{h}}_{m}=\mathbf{h}_{m}-\mathbf{h}_{\mathcal{Z}}$.

	\end{IEEEproof}

	According to Theorem~\ref{theorem1-1},  there is $J_{\mathrm{LMMSE}}\approx J(\mathbf{f}_{\boldsymbol{\theta}_{\mathcal{Z}}})$ as the size of the training sample set gets sufficiently large, and we reach the following conclusion for the linear systems as
	\begin{align}\label{comparison}
	J_{\mathrm{LMMSE}}\approx J(\mathbf{f}_{\boldsymbol{\theta}_{\mathcal{Z}}})\leq J_{\mathrm{LS}}.
	\end{align}
	
	\begin{remark}\label{remark2}
		\begin{itemize}
			
			\item The affine transmission $\mathcal{A}$ can be represented by a ReLU DNN with more than two layers, which can be expressed as
			\begin{align}\label{affine-remark}
			\mathcal{A}=(\mathbf{I}_{d}\circ\cdots\circ\varphi_{d}\circ\mathbf{I}_{d}\circ\varphi_{d}\circ\mathcal{A})+(-\mathbf{I}_{d}\circ\cdots\circ\varphi_{d}\circ\mathbf{I}_{d}\circ\varphi_{d}\circ(-\mathcal{A})).
			\end{align}
			Setting $\mathcal{A}$ in~\eqref{affine-remark} as the identity transformation, we can replicate the output of any hidden layer of a ReLU DNN by adding one or multiple of hidden layers. Moreover, the width of  the ReLU DNN can be arbitrarily large provided that its size is bigger than $2d$. Therefore, Theorem~\ref{theorem1-1} is extensible to a wide class of the DL estimators. 
			\item The estimated channel derived in practice through numerical optimization is only effective for a small range of the input space that contains the training samples, though $\mathbf{f}_{\boldsymbol{\theta}_{\mathcal{Z}}}(\mathbf{x})$ in~\eqref{theo2-1} is defined at a global scope. This phenomenon will be discussed in Section~\ref{incomplete}.
		\end{itemize}
	\end{remark}

\subsubsection{Nonlinear Systems}
	
Wireless communication systems often suffer from some nonlinear effects, e.g., imperfect power amplifier (PA)~\cite{costa1999impact,costa2002m} and quantization error of analog to digital converter (ADC)~\cite{XULIANG2019}. It is hard to establish precise signal models for most nonlinear systems and designing the corresponding channel estimation is very challenging. In general, the theoretical framework in nonlinear communication systems is captured by the following statistical model
\begin{align}\label{non}
\mathbf{x}_{0}=\mathbf{f}_{\mathrm{NL}}(\tau\mathbf{h}+\mathbf{n}),
\end{align}
where $\mathbf{f}_{\mathrm{NL}}(\cdot)$ denotes the nonlinear distortion imposed on the received signal. The optimal estimate of $\mathbf{h}$ is still provided by $\mathbf{h}_{\mathrm{MMSE}}$ in~\eqref{mmse}. Unlike the linear systems, the explicit form of $\mathbf{h}_{\mathrm{MMSE}}$ is impossible to obtain in most nonlinear cases. However, the LS and LMMSE estimators are still used for channel estimation in nonlinear systems before the advent of the DL estimator. There exit two major problems that affect the performance of the LS and LMMSE estimators. First, it is difficult to acquire accurate channel statistics due to complexity and uncertainty of the nonlinear systems. Second, the system error of fitting nonlinear objective function with a linear estimator is big. Either way, the LS and LMMSE estimator are not good options for channel estimation in the nonlinear systems. 

On the other hand, the DL estimator is far more expressive than the LMMSE estimator at function representation due to the piecewise linear property and is more suitable to deploy in the nonlinear systems. The estimated channel of the DL estimator based on the ReLU DNN can arbitrarily well approximate to $\mathbf{h}_{\mathrm{MMSE}}$  as long as the structure of the DNN is suitably configured and the size of the dataset $\mathcal{Z}$ is sufficiently large. 

From~\eqref{expected}, $J(\mathbf{f}_{\boldsymbol{\theta}_{\mathcal{Z}}})$ can break into two different terms as
\begin{align}\label{decom}
J(\mathbf{f}_{\boldsymbol{\theta}_{\mathcal{Z}}})=J(\mathbf{f}_{\boldsymbol{\theta}_{o}})+[J(\mathbf{f}_{\boldsymbol{\theta}_{\mathcal{Z}}})-J(\mathbf{f}_{\boldsymbol{\theta}_{o}})].
\end{align}

The second term, $J(\mathbf{f}_{\boldsymbol{\theta}_{\mathcal{Z}}})-J(\mathbf{f}_{\boldsymbol{\theta}_{o}})$, in the RHS of~\eqref{decom} is non-negative and is determined by the dataset $\mathcal{Z}$ when the structure of the DNN is fixed, which is called the generalization error.  

The first term $J(\mathbf{f}_{\boldsymbol{\theta}_{o}})$ in~\eqref{decom} is determined by the structure of the DNN and is independent of sampling set. From~\eqref{optimal}, we can further decompose it as
\begin{align}\label{decomp1}
J(\mathbf{f}_{\boldsymbol{\theta}_{o}})=\mathbb{E}\{\|\mathbf{f}_{\boldsymbol{\theta}_{o}}(\mathbf{x})-\mathbf{f}_{o}(\mathbf{x})\|^{2}_{2}\}+J(\mathbf{f}_{o}).
\end{align}
From~\eqref{decomp1}, $J(\mathbf{f}_{\boldsymbol{\theta}_{o}})$  is determined by $\mathbb{E}\{\|\mathbf{f}_{\boldsymbol{\theta}_{o}}(\mathbf{x})-\mathbf{f}_{o}(\mathbf{x})\|^{2}_{2}\}$, which is referred to as the approximation error.

We will analyze the performance of the DL estimator through quantifying the approximation error and generalization error of the DL estimator in the nonlinear systems.
	
First, the approximation error in~\eqref{decom} of the DL estimator can be narrowed down with any precision by a ReLU DNN of finite length, as demonstrated by the following theorem.

	\begin{theorem}\label{theorem1}
		For any $\varepsilon>0$, there exits an optimized DL estimator $\mathbf{f}_{\boldsymbol{\theta}_{o}}(\mathbf{x})$ built on the ReLU DNN with at most $	
		\lceil\log_{2}(d+1)\rceil
		$ hidden layers such that
		\begin{align}
		\mathbb{E}\{\|\mathbf{f}_{\boldsymbol{\theta}_{o}}(\mathbf{x})-\mathbf{f}_{o}(\mathbf{x})\|^{2}_{2}\}\leq\varepsilon.
		\end{align}
	\end{theorem}
	\begin{IEEEproof}
	From~\eqref{network}, $\mathbf{f}_{\boldsymbol{\theta}_{o}}(\mathbf{x})$ is equivalent to a $\mathbb{R}^{d}\rightarrow\mathbb{R}^{d}$ piecewise linear function. According to~\cite[Theorem 2.1]{arora2018understanding}, any $\mathbb{R}^{d}\rightarrow\mathbb{R}$ piecewise linear function can be represented by a DL estimator that is built on a ReLU DNN with no more than $\lceil\log_{2}(d+1)\rceil$ hidden layers. 
	
	On the other hand, any function in $\ell_{2}$ can be approximated by a piecewise linear function with arbitrary precision~\cite{royden2010real}. Let $f_{o,i}(\mathbf{x})$ be the $i$-th entry of $\mathbf{f}_{o}(\mathbf{x})$ for $i\in\{1,\ldots,d\}$ and there exists $d$ $\mathbb{R}^{d}\rightarrow\mathbb{R}$ piecewise linear functions to approximate $\{f_{o,1}(\mathbf{x}),\ldots,f_{o,d}(\mathbf{x})\}$ with arbitrary precision. Such a set of $d$ piecewise linear functions can be represented by $d$ ReLU DNNs each with at most $\lceil\log_{2}(d+1)\rceil$ hidden layers. We can simply put these ReLU DNNs in parallel and combine their outputs to compose a single ReLU DNN. 
	
	The depths of these ReLU DNNs may be different and we need to align their depths for the composition. Since the output of any hidden layer of a ReLU DNN can be replicated by adding hidden layers, we can simply add one or multiple of hidden layers for each ReLU DNN to align the depths of these ReLU DNNs. Let $l_{\mathrm{max}}$ be the maximum depth of these ReLU DNNs and then the aligned depth is just given by $l_{\mathrm{max}}$, which is upper bounded by $\lceil\log_{2}(d+1)\rceil$.  Therefore, there exits a DL estimator with parameter set $\boldsymbol{\theta}_{\varepsilon}$  with at most 
	$\lceil\log_{2}(d+1)\rceil$ hidden layers
	such that
	\begin{align}
	\|\mathbf{f}_{\boldsymbol{\theta}_\varepsilon}(\mathbf{x})-\mathbf{f}_{o}(\mathbf{x})\|^{2}_{2}\leq\varepsilon
	\end{align}
	for any $\varepsilon>0$.
	From~\eqref{decomp1}, $\mathbf{f}_{\boldsymbol{\theta}_{o}}(\mathbf{x})$ is the best approximation to $\mathbf{f}_{o}(\mathbf{x})$ over all the DL estimators, and we have
	\begin{align}\label{theo1-2}
	\|\mathbf{f}_{\boldsymbol{\theta}_{o}}(\mathbf{x})-\mathbf{f}_{o}(\mathbf{x})\|_{2}^{2}\leq\|\mathbf{f}_{\boldsymbol{\theta}_\varepsilon}(\mathbf{x})-\mathbf{f}_{o}(\mathbf{x})\|_{2}^{2}\leq\varepsilon.
	\end{align}
	From~\eqref{theo1-2}, it holds that
	\begin{align}
	\mathbb{E}\{\|\mathbf{f}_{\boldsymbol{\theta}_{o}}(\mathbf{x})-\mathbf{f}_{o}(\mathbf{x})\|^{2}_{2}\}\leq\mathbb{E}\{\varepsilon\}=\varepsilon,
	\end{align}
	which completes the proof.
	\end{IEEEproof}
\begin{remark}\label{remark3}
	\begin{itemize}
		\item
	Theorem~\ref{theorem1} shows that the approximation error in~\eqref{decomp1} can be bounded with arbitrary precision if the underlying ReLU DNN of the DL estimator is reasonably configured. 
	\item Theorem~\ref{theorem1} also indicates that the DL estimator is powerful at function representation and does not restrict to any type of signal models or channel statistics. If no specific models are known a priori or complicated nonlinear systems are presented, the DL estimator will be a preferred choice for channel estimation.
	\end{itemize}
	\end{remark}

	The following theorem demonstrates that  the generalization error in~\eqref{decom} asymptotically approaches to zero as $|\mathcal{Z}|$ increases.

	\begin{theorem}\label{theorem2}
		Suppose that both $\|\boldsymbol{\theta}_{o}\|_{2}$ and $\|\boldsymbol{\theta}_{\mathcal{Z}}\|_{2}$ are finite. Then, it holds that
		\begin{align}
		J(\mathbf{f}_{\boldsymbol{\theta}_{\mathcal{Z}}})-J(\mathbf{f}_{\boldsymbol{\theta}_{o}})\stackrel{\mathbf{P}}{\longrightarrow}0.
		\end{align}
	\end{theorem}
	\begin{IEEEproof}
		For finite $\|\boldsymbol{\theta}_{o}\|_{2}$ and $\|\boldsymbol{\theta}_{\mathcal{Z}}\|_{2}$, we have 
		\begin{align}
		J_{\mathcal{Z}}(\mathbf{f}_{\boldsymbol{\theta}_{o}})-J(\mathbf{f}_{\boldsymbol{\theta}_{o}})\stackrel{\mathbf{P}}{\longrightarrow}0\ \mathrm{and}\ 
		J_{\mathcal{Z}}(\mathbf{f}_{\boldsymbol{\theta}_{\mathcal{Z}}})-J(\mathbf{f}_{\boldsymbol{\theta}_{\mathcal{Z}}})\stackrel{\mathbf{P}}{\longrightarrow}0
		\end{align} 
		according to the law of large numbers~\cite{mendenhall2012introduction}.
		Then, the following inequalities hold 
		\begin{align}
		0&\leq J(\mathbf{f}_{\boldsymbol{\theta}_{\mathcal{Z}}})-J(\mathbf{f}_{\boldsymbol{\theta}_{o}})\nonumber\\
		&=[J(\mathbf{f}_{\boldsymbol{\theta}_{\mathcal{Z}}})-J(\mathbf{f}_{\boldsymbol{\theta}_{o}})]-[J_{\mathcal{Z}}(\mathbf{f}_{\boldsymbol{\theta}_{\mathcal{Z}}})-J_{\mathcal{Z}}(\mathbf{f}_{\boldsymbol{\theta}_{o}})]+[J_{\mathcal{Z}}(\mathbf{f}_{\boldsymbol{\theta}_{\mathcal{Z}}})-J_{\mathcal{Z}}(\mathbf{f}_{\boldsymbol{\theta}_{o}})]\nonumber\\
		&\leq [J(\mathbf{f}_{\boldsymbol{\theta}_{\mathcal{Z}}})-J(\mathbf{f}_{\boldsymbol{\theta}_{o}})]-[J_{\mathcal{Z}}(\mathbf{f}_{\boldsymbol{\theta}_{\mathcal{Z}}})-J_{\mathcal{Z}}(\mathbf{f}_{\boldsymbol{\theta}_{o}})]\stackrel{\mathbf{P}}{\longrightarrow}0.
		\end{align}
		Therefore, $J(\mathbf{f}_{\boldsymbol{\theta}_{\mathcal{Z}}})-J(\mathbf{f}_{\boldsymbol{\theta}_{o}})\stackrel{\mathbf{P}}{\longrightarrow}0$.
	\end{IEEEproof}

	Together with Theorem~\ref{theorem1}, the following corollary presents our main conclusion on the performance of the DL estimator in the nonlinear systems.
	\begin{corollary}\label{corollary}
		For finite $\|\boldsymbol{\theta}_{o}\|_{2}$ and $\|\boldsymbol{\theta}_{\mathcal{Z}}\|_{2}$ and any $\varepsilon>0$, there exits a DL estimator powered by the ReLU DNN with at most
		$\lceil\log_{2}(d+1)\rceil
	$ hidden layers
		such that
		\begin{align}\label{coro1-1}
		\lim_{|\mathcal{Z}|\rightarrow+\infty}\mathbf{P}\big([J(\mathbf{f}_{\boldsymbol{\theta}_{\mathcal{Z}}})-J(\mathbf{f}_{o})]>\varepsilon\big)=0
		\end{align}
		for any $\varepsilon>0$.
	\end{corollary}
	\begin{IEEEproof}
	According to~\eqref{decom} and~\eqref{decomp1}, $J(\mathbf{f}_{\boldsymbol{\theta}_{\mathcal{Z}}})-J(\mathbf{f}_{o})$ is decomposed into 
	\begin{align}
	J(\mathbf{f}_{\boldsymbol{\theta}_{\mathcal{Z}}})-J(\mathbf{f}_{o})=\underbrace{\mathbb{E}\{\|\mathbf{f}_{\boldsymbol{\theta}_{o}}(\mathbf{x})-\mathbf{f}_{o}(\mathbf{x})\|^{2}_{2}\}}_{\mathrm{Approximation\ error}}+\underbrace{J(\mathbf{f}_{\boldsymbol{\theta}_{\mathcal{Z}}})-J(\mathbf{f}_{\boldsymbol{\theta}_{o}})}_{\mathrm{Generalization\ error}}.
	\end{align}
	
	From Theorem~\ref{theorem1}, there exits an optimized DL estimator $\mathbf{f}_{\boldsymbol{\theta}_{o}}(\mathbf{x})$ with
	at most $
\lceil\log_{2}(d+1)\rceil
	$ hidden layers such that $
	\mathbb{E}\{\|\mathbf{f}_{\boldsymbol{\theta}_{o}}(\mathbf{x})-\mathbf{f}_{o}(\mathbf{x})\|^{2}_{2}\}\leq\varepsilon$
	for any $\varepsilon>0$. From Theorem~\ref{theorem2}, $J(\mathbf{f}_{\boldsymbol{\theta}_{\mathcal{Z}}})-J(\mathbf{f}_{\boldsymbol{\theta}_{o}})\stackrel{\mathbf{P}}{\longrightarrow}0$. Combining Theorems~\ref{theorem1} and Theorem~\ref{theorem2}, we have
	\begin{align}
	\lim_{|\mathcal{Z}|\rightarrow+\infty}\mathbf{P}\big([J(\mathbf{f}_{\boldsymbol{\theta}_{\mathcal{Z}}})-J(\mathbf{f}_{o})]>\varepsilon\big)=0
	\end{align}
	for any $\varepsilon>0$, which completes the proof.
	\end{IEEEproof}
\begin{figure}[t]
	\vskip 0.2in
	\centering
	
	\includegraphics[width=0.8\columnwidth]{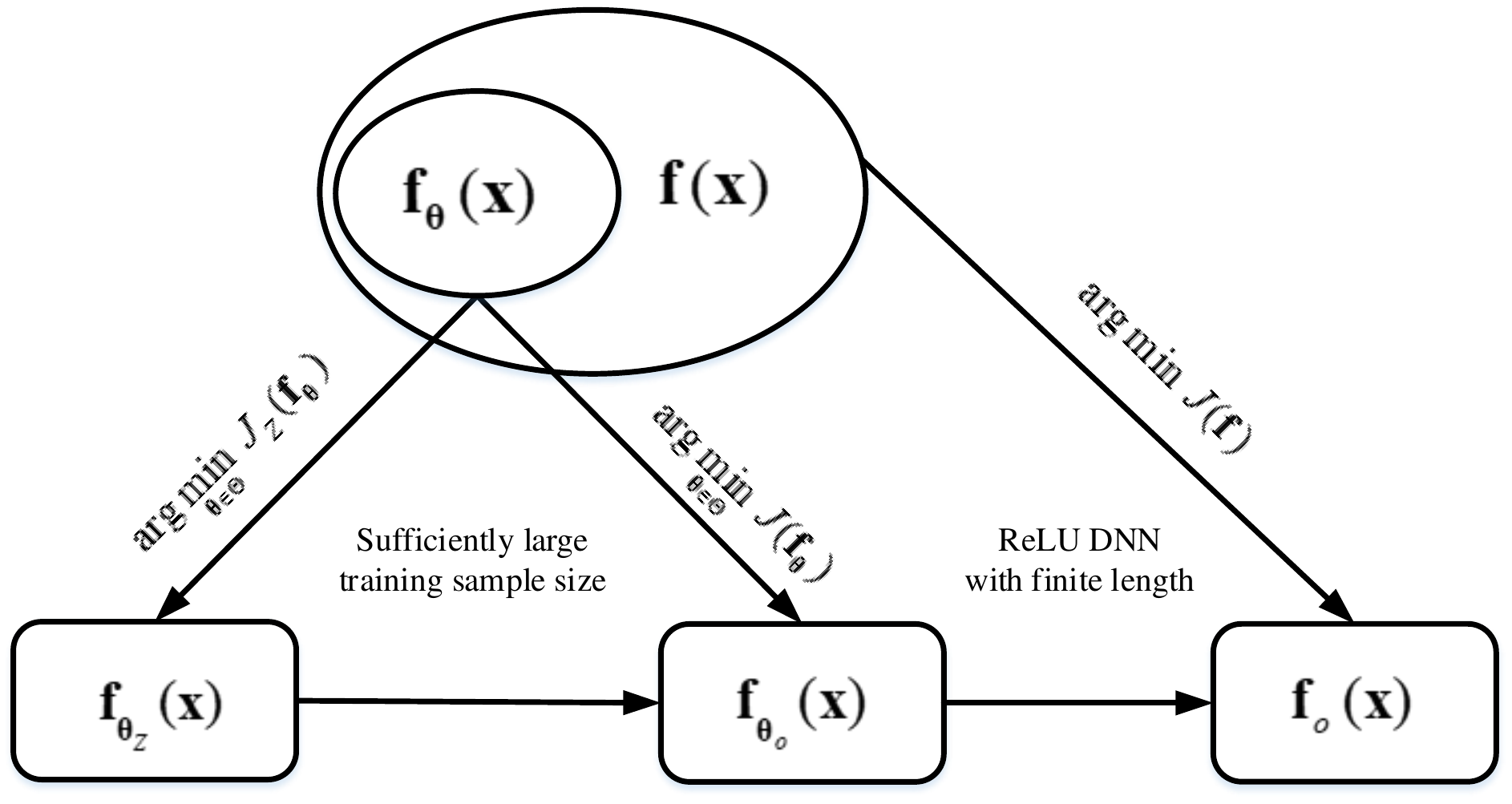}


	\caption{The relationship between $\mathbf{f}_{\boldsymbol{\theta}_{\mathcal{Z}}}(\mathbf{x})$, $\mathbf{f}_{\boldsymbol{\theta}_{o}}(\mathbf{x})$, and $\mathbf{f}_{o}(\mathbf{x})$.}
	\label{figure1-1}
	\vskip -0.2in
\end{figure}

	Fig.~\ref{figure1-1} illustrates the relationship between $\mathbf{f}_{\boldsymbol{\theta}_{\mathcal{Z}}}(\mathbf{x})$, $\mathbf{f}_{\boldsymbol{\theta}_{o}}(\mathbf{x})$, and $\mathbf{f}_{o}(\mathbf{x})$ to better understand Theorem~\ref{theorem1} and Theorem~\ref{theorem2}. These two theorems demonstrate that the estimated channel of the DL estimator based on the ReLU DNN can arbitrarily well-approximate to the estimate of the MMSE estimator, i.e., $\mathbf{f}_{o}(\mathbf{x})$ or $\mathbf{h}_{\mathrm{MMSE}}$, as the size of the training sample set, $|\mathcal{Z}|$, gets large. We then derive the main result on the performance of the DL estimator based on Corollary~\ref{corollary} as 
	\begin{align}\label{con}
	J(\mathbf{f}_{o})\approx J(\mathbf{f}_{\boldsymbol{\theta}_{\mathcal{Z}}}).
	\end{align}
	
	From Remark~\ref{remark2}, any affine transformation $\mathcal{A}$ is representable by the ReLU DNN when the number of layers is more than $2$ and the size is larger than $2d$. If there exits a DL estimator satisfying Corollary~\ref{corollary}, we can simply add hidden layers and neurons at each hidden layer to make sure that the underlying network structure of the DL estimator can represent the affine transformation $\mathcal{A}$. Then, from~\eqref{app1-2} and~\eqref{affine1}, we have
	\begin{align}
	J(\mathbf{f}_{\boldsymbol{\theta}_{\mathcal{Z}}})\leq J(\mathcal{A})
	\end{align}
	when $|\mathcal{Z}|$ is sufficiently large. Replacing $\mathcal{A}$ with $\mathbf{h}_{\mathrm{LS}}$ or $\mathbf{h}_{\mathrm{LMMSE}}$, we can further write~\eqref{con} as
	\begin{align}
	J(\mathbf{f}_{o})\approx J(\mathbf{f}_{\boldsymbol{\theta}_{\mathcal{Z}}})\leq J_{\mathrm{LS}},J_{\mathrm{LMMSE}}.
	\end{align}

	\begin{remark}
	\begin{itemize}
		\item
	Corollary~\ref{corollary} provides a theoretical insight into the excellent performance of the DL estimator for channel estimation under imperfect environments,  where the performance of the LS and LMMSE estimators degrades significantly. In this case, the DL estimator is still able to build up a stable and precise model to estimate $\mathbf{h}$ due to its universal approximation illustrated by Corollary~\ref{corollary}. Therefore, the DL estimator has a great potential to combat nonlinear distortion and some other unknown detrimental effects in real world communication systems.

	\item

	Owing to no assumption about underlying signal model, the DL estimator has to take sufficiently large training data to train an effective estimator from scratch which is relatively inefficient compared to the LS or LMMSE estimators. In fact, we can retrain a learned DL estimator that is originally trained at similar scenarios to accelerate the training process as what the transfer learning has done in image processing~\cite{pan2009survey}.
\end{itemize}
 \end{remark}

	\section{Robustness to Mismatched Information}\label{incomplete}
	
	The optimality of the LMMSE and DL estimators depends on the perfect knowledge of channel statistics and matching training data, respectively, while it is a typical problem that the channel covariance matrix $\mathbf{R}$ is not perfectly known or the statistics of training data do not match the deployed environments. In this section, we analyze channel estimation with inaccurate channel statistics and mismatched training data for the linear system model in~\eqref{r_train} and show how these imperfections affect the performance of the LMMSE and DL estimators.
	
	\subsection{LMMSE Estimator}
	Denote the channel covariance matrix used by the LMMSE estimator as $\mathbf{R}_{1}=\mathbf{R}+\mathbf{\Omega}$, where $\mathbf{\Omega}$ is the $d\times d$ Hermitian random error matrix independent of $\mathbf{R}$. Replacing $\mathbf{R}$ by $\mathbf{R}_{1}$ in \eqref{mmse1}, the LMMSE estimator under inaccurate channel statistic can be expressed as
	\begin{align}\label{estimate}
	\mathbf{h}_{\mathrm{LM-ER}}=\mathbf{R}_{1}(\mathbf{R}_{1}+\sigma_{n}^{2}\mathbf{I}_{d})^{-1}\mathbf{x},
	\end{align}
	and the corresponding MSE is given by
	\begin{align}\label{lmmse}
	J_{\mathrm{LM-ER}}=\mathrm{tr}\bigg\{\left(\mathbf{R}_{1}^{-1}+\frac{1}{\sigma_{n}^{2}}\mathbf{I}_{d}\right)^{-1}-\mathbf{\Pi}\mathbf{\Omega}\mathbf{\Pi}^{T}\bigg\},
	\end{align}
	where $\mathbf{\Pi}=\mathbf{I}_{d}-\mathbf{R}_{1}(\mathbf{R}_{1}+\sigma_{n}^{2}\mathbf{I}_{d})^{-1}$. 
	
	It is difficult to figure out how $\mathbf{\Omega}$ affects the estimation accuracy of the LMMSE estimator directly from~\eqref{lmmse}. However, we can take the uncorrelated channel as an example to demonstrate the influence of $\mathbf{\Omega}$ on $J_{\mathrm{LM-ER}}$ in general since the channels between different received antennas are asymptotically uncorrelated when $d$ gets large~\cite{rusek2013scaling}

	Suppose that the covariance matrix of $\mathbf{h}$ is diagonal with $\mathbf{R}=\sigma^{2}\mathbf{I}_{d}$, where $\sigma^{2}$ is element-wise variance. Moreover, assume that $\mathbf{\Omega}$ can be decomposed into
	\begin{align}\label{decompose}
	\mathbf{\Omega}=\mathbf{U}\mathbf{\Sigma}\mathbf{U}^{T},
	\end{align}
	where $\mathbf{U}$ is the $d\times d$ eigenvector matrix and $\mathbf{\Sigma}$ is the $d\times d$ eigenvalue matrix. Substituting $\mathbf{R}=\sigma^{2}\mathbf{I}_{d}$ and~\eqref{decompose} into~\eqref{estimate}, we can rewrite $J_{\mathrm{LM-ER}}$ as
	\begin{align}\label{lmmse1}
	J_{\mathrm{LM-ER}}=J_{\mathrm{LMMSE}}+\sum_{i=1}^{d}\frac{ \sigma_{e,i}^{4}\sigma_{n}^{4}d}{(\sigma^{2}+\sigma_{e,i}^{2}+\sigma_{n}^{2})^{2}(\sigma^{2}+\sigma_{n}^{2})},
	\end{align}
	where $\sigma_{e,i}^{2}$ is the $i$-th diagonal element of $\mathbf{\Sigma}$. 
	
	Reversely, if $\mathbf{R}=\mathbf{R}_{1}+\mathbf{\Omega}$ and $\mathbf{R}=\sigma^{2}\mathbf{I}_{d}$, then we can still obtain the same form of $J_{\mathrm{LM-ER}}$ in~\eqref{lmmse1}.
	
	\begin{remark}
		\begin{itemize}
			\item
		According to~\eqref{lmmse1}, $J_{\mathrm{LM-ER}}$ is always larger than $J_{\mathrm{LMMSE}}$ and is increased with $\sigma_{e,i}^{2}$. The performance of the LMMSE estimator is mainly determined by the accuracy of $\mathbf{R}$. 
		\item The DL estimator needs to know neither the exact signal model nor the information of $\mathbf{R}$ to estimate $\mathbf{h}$. Whether $\mathbf{R}$ is accurate or not does not affect the accuracy of the DL estimator. Hence, the DL estimator will outperform the LMMSE estimator if $\sigma_{e,i}^{2}$ exceeds certain threshold, i.e., $J_{\mathrm{LM-ER}}\geq J(\mathbf{f}_{\boldsymbol{\theta}_{\mathcal{Z}}})$.
		\end{itemize}
	\end{remark}
	
	\subsection{DL Estimator}\label{dlerror}
	
	The DL estimator is data-driven with its performance mainly determined by how well the training data matches the working environment and mismatched training data will lead to significant performance degradation. Furthermore, different from the LS and LMMSE estimators defined w.r.t. the whole input space, the learned DL estimator can only make valid estimated channels for the inputs that are restricted to the regions where training samples are not empty and will behave randomly outside these regions, as discussed in Remark~\ref{remark2}. The restricted effective input range puts severe limit on the performance of the DL estimator.
	
	Let 
	\begin{align}
	\mathcal{Z}_{k}=\{m\,|\,\mathbf{x}_{m}\in\mathcal{X}_{k},m= 1,\ldots,|\mathcal{Z}|\}
	\end{align}
	be the set of index of samples that fall into $\mathcal{X}_{k}$. Note that $(|\mathcal{Z}_{1}|,\ldots,|\mathcal{Z}_{K}|)$ is an i.i.d. multinomial random variable with probability $(\psi(\mathcal{X}_{1}),\ldots,\psi(\mathcal{X}_{K}))$ and satisfies $\sum_{i=1}^{K}|\mathcal{Z}_{i}|=|\mathcal{Z}|$. The following theorem provides the explicit form of $\mathbf{f}_{\boldsymbol{\theta}_{\mathcal{Z}}}(\mathbf{x})$ in the linear systems with sufficiently large $|\mathcal{Z}|$. Using~\eqref{network}, we rewrite the empirical loss as
	\begin{align}\label{empirical1}
	J_{\mathcal{Z}}(\mathbf{f}_{\boldsymbol{\theta}_{\mathcal{Z}}})=\frac{1}{|\mathcal{Z}|}\sum_{k=1}^{K}\sum_{m\in\mathcal{Z} _{k}}\mathrm{tr}\big\{(\mathbf{h}_{m}-\mathbf{W}_{\mathcal{X}_{k}}\mathbf{x}_{m}-\mathbf{b}_{\mathcal{X}_{k}})
	(\mathbf{h}_{m}-\mathbf{W}_{\mathcal{X}_{k}}\mathbf{x}_{m}-\mathbf{b}_{\mathcal{X}_{k}})^{T}\big\}.
	\end{align}
	
	An important issue is that only a small number of partitioned regions within $\mathcal{X}$, where $\mathbf{x}$ falls into with high probabilities, contain training samples. For the regions without training samples, i.e., $|\mathcal{Z}_{k}|=0$, the DL estimator is unable to optimize its estimate through $J_{\mathcal{Z}}(\mathbf{f}_{\boldsymbol{\theta}_{\mathcal{Z}}})$, as shown in~\eqref{empirical1}, and will simply output a random estimated channel if $\mathbf{x}$ is located at these regions. In general, this limitation has a little impact on the performance of the DL estimator when training data has accurate statistics since the probability that $\mathbf{x}$ falls into the regions without training samples is very low. However, if the statistics of training data do not match real channels, such a probability can not be ignored and the limitation on the effective input range will lead to serious issues.

	Denote by $\mathbf{h}_{\mathrm{DL-ER}}$ and  $J_{\mathrm{DL-ER}}$ the estimate and the MSE of the DL estimator, respectively, when training data mismatches the real channel. The MSE of the DL estimator under mismatched training data is discussed in the following two separate cases.
	
	\subsubsection{Case I} Assume that $\mathbf{h}_{\mathrm{er}}$ distributes in a broader range than $\mathbf{h}$, where the variance of $\mathbf{h}_{\mathrm{er}}$ is larger than that of  $\mathbf{h}$. The corresponding statistical models of the training data are described as
	\begin{align}\label{statistical3}
	\mathbf{h}_{\mathrm{er}}=\mathbf{h}+\boldsymbol{\zeta},
	\end{align}
	and
	\begin{align}\label{statistical1}
	\mathbf{x}_{\mathrm{er}}=\tau\mathbf{h}_{\mathrm{er}}+\mathbf{n},
	\end{align}
	where $\mathbf{h}$ is Gaussian distributed and $\boldsymbol{\zeta}$ denotes the $d\times 1$ zero mean random error vector that is independent of $\mathbf{h}$ with covariance matrix $\mathbf{\Omega}_{\zeta}=\mathbb{E}\{\boldsymbol{\zeta}\boldsymbol{\zeta}^{T}\}$. 
	
	In Case $1$, the probability that $\mathbf{x}$ falls into the regions without training samples is still close to zero since $\mathbf{x}_{\mathrm{er}}$ is more broadly distributed than $\mathbf{x}$. From Theorem~\ref{theorem1-1}, the target estimator that the DL estimator approaches to as $|\mathcal{Z}|$ gets large is the MMSE estimator w.r.t. $\mathbf{h}_{\mathrm{er}}$, and the corresponding estimated channel is given by  
	\begin{align}\label{error}
	\mathbf{h}_{\mathrm{MM-ER}}=\mathbf{C}_{\mathbf{h}_{\mathrm{er}}\mathbf{x}_{\mathrm{er}}}\mathbf{C}_{\mathbf{x}_{\mathrm{er}}\mathbf{x}_{\mathrm{er}}}^{-1}\mathbf{x},
	\end{align}
	where $\mathbf{C}_{\mathbf{h}_{\mathrm{er}}\mathbf{x}_{\mathrm{er}}}$ is the cross-covariance of $\mathbf{h}_{\mathrm{er}}$ and $\mathbf{x}_{\mathrm{er}}$ and $\mathbf{C}_{\mathbf{x}_{\mathrm{er}}\mathbf{x}_{\mathrm{er}}}$ is the covariance of $\mathbf{x}_{\mathrm{er}}$. If the DL estimator is properly configured and $|\mathcal{Z}|$ is sufficiently large, then we have
	\begin{align}  
	\mathbf{h}_{\mathrm{DL-ER}}\approx\mathbf{h}_{\mathrm{MM-ER}}
	\end{align} 
	according to Theorem~\ref{theorem1-1}. From~\eqref{optimal}, the corresponding MSE is
	\begin{align}\label{dlerr}
	J_{\mathrm{DL-ER}}\approx J_{\mathrm{LMMSE}}+\|(\mathbf{C}_{\mathbf{h}_{\mathrm{er}}\mathbf{x}_{\mathrm{er}}}\mathbf{C}_{\mathbf{x}_{\mathrm{er}}\mathbf{x}_{\mathrm{er}}}^{-1}-\mathbf{C}_{\mathbf{h}\mathbf{x}}\mathbf{C}_{\mathbf{x}\mathbf{x}}^{-1})\mathbf{x}\|_{2}^{2},
	\end{align}
	where $\mathbf{C}_{\mathbf{x}\mathbf{x}}$ is the covariance of $\mathbf{x}$.
	
	Similar to $J_{\mathrm{LM-ER}}$ in~\eqref{lmmse1}, how $\boldsymbol{\zeta}$ affects $J_{\mathrm{DL-ER}}$ is difficult to justify from~\eqref{dlerr}. To provide some insight into the influence of the mismatched training data on the DL estimator, we assume that $\mathbf{R}=\sigma^{2}\mathbf{I}_{d}$. Then, the covariance matrix $\mathbf{\Omega}_{\zeta
	}$ can be decomposed into
	\begin{align}\label{decompose1}
	\mathbf{\Omega}_{\zeta}=\mathbf{U}_{\zeta}\mathbf{\Sigma}_{\zeta}\mathbf{U}_{\zeta}^{T},
	\end{align}
	where $\mathbf{U}_{\zeta}$ is the $d\times d$ eigenvector matrix and $\mathbf{\Sigma}_{\zeta}$ is the $d\times d$ eigenvalue matrix. Substituting $\mathbf{R}=\sigma^{2}\mathbf{I}_{d}$ and~\eqref{decompose1} into~\eqref{dlerr} yields
	\begin{align}\label{DL-ER}
	J_{\mathrm{DL-ER}}\approx J_{\mathrm{LMMSE}}+\sum_{i=1}^{d}\frac{\sigma_{\zeta,i}^{4}\sigma_{n}^{4}}{(\sigma^{2}+\sigma_{\zeta,i}^{2}+\sigma_{n}^{2})^{2}(\sigma^{2}+\sigma_{n}^{2})},
	\end{align}
	where $\sigma_{\zeta,i}^{2}$ is the $i$-th diagonal element of $\mathbf{\Sigma}_{\zeta}$ and quantifies the mismatch degree between the training data and the real systems. The obtained $J_{\mathrm{DL-ER}}$ in~\eqref{DL-ER} is similar to $J_{\mathrm{LM-ER}}$ in~\eqref{lmmse1} and also increases with $\sigma_{\zeta,i}^{2}$.
	
	\subsubsection{Case II} We consider that the input-output pair of training data is generated from the following statistical model
	\begin{align}\label{statistical5}
	\mathbf{h}=\mathbf{h}_{\mathrm{er}}+\boldsymbol{\zeta},
	\end{align}
	and
	\begin{align}\label{statistical6}
	\mathbf{x}_{\mathrm{er}}=\tau\mathbf{h}_{\mathrm{er}}+\mathbf{n}.
	\end{align}
	
	In Case II, $\mathbf{x}$ distributes in a broader range than $\mathbf{x}_{\mathrm{er}}$, and the probability that $\mathbf{x}$ falls at regions without training samples is much higher than Case I. From~\eqref{empirical1}, the DL estimator is not optimized for the whole input space, and its effective input range is dependent on the training data distribution. The estimated channels of the DL estimator corresponding to the inputs at empty regions are totally random and unacceptable if the discrepancy between $\mathbf{h}$ and $\mathbf{h}_{\mathrm{er}}$ is very large. In this case, the DL estimator basically fails to provide a reliable estimated channel, and its performance degrades severely.
	
	\begin{remark}
		\begin{itemize}
			\item
		In Case I, the probabilities that the inputs are located at the regions without training samples are negligible. The DL estimator can well approximate $\mathbf{h}_{\mathrm{MM-ER}}$ and provide a stable estimated channel. Hence, the limited effective input range has a little impact on the performance of the DL estimator.
		\item
		In Case II, the probabilities that the inputs are located at the regions without training samples can not be neglected. The limitation on the effective input range of the DL estimator gets really serious, and the DL estimator is unable to provide a valid estimated channel when the input is located outside the regions with training samples. The LMMSE estimator, however, is designed over the whole input space based on the expert knowledge, and therefore the error introduced by the discrepancy between $\mathbf{R}$ and $\mathbf{R}_{1}$ is controllable no matter how $\mathbf{\Omega}$ varies. In this case, the traditional LMMSE estimator is more robust to the imperfect data than the DL estimator.
		\end{itemize}
	\end{remark}

	\section{Simulation Results}\label{simulation}
	
	In this section, computer simulation is conducted to provide further evidence and insights into the performance assessment of various estimators, which also verifies the advantages and disadvantages of the DL channel estimation.

	\subsection{Linear Systems}\label{linear-simu}
	
	\begin{figure}[!t]
		\centering
		\includegraphics[width=0.75\columnwidth]{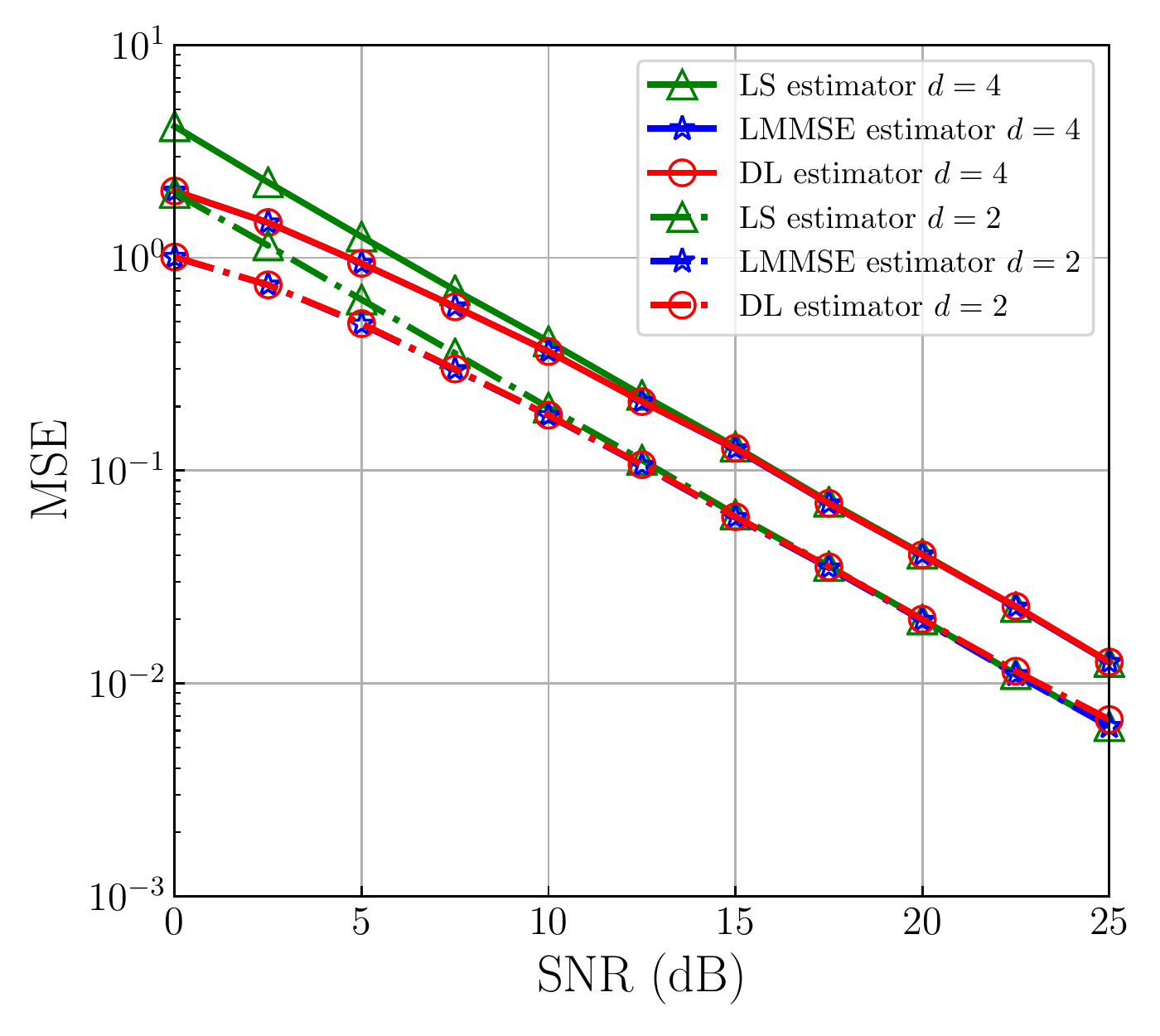}
		\caption{The MSE performance of the LS, LMMSE, and DL estimators versus SNR under linear signal model.}
		\label{figure2}
	\end{figure}
	
		\begin{figure}[!t]
		\centering
		\includegraphics[width=0.75\columnwidth]{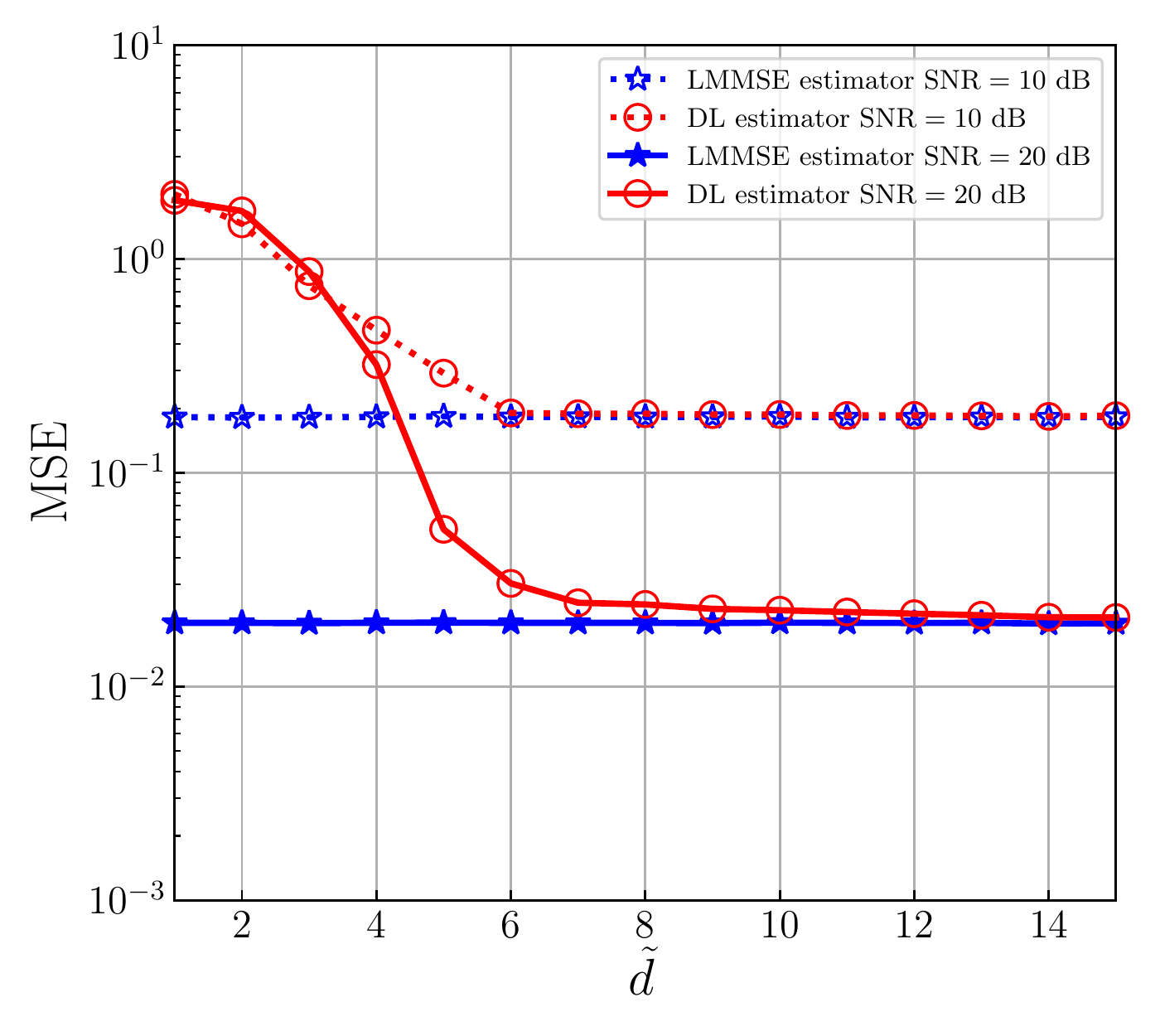}
		\caption{The MSE performance of the DL estimator versus $\tilde{d}$ under linear signal model.}
		\label{figure2-c}
	\end{figure}
	
	\begin{figure}[!t]
		\centering
		\includegraphics[width=0.75\columnwidth]{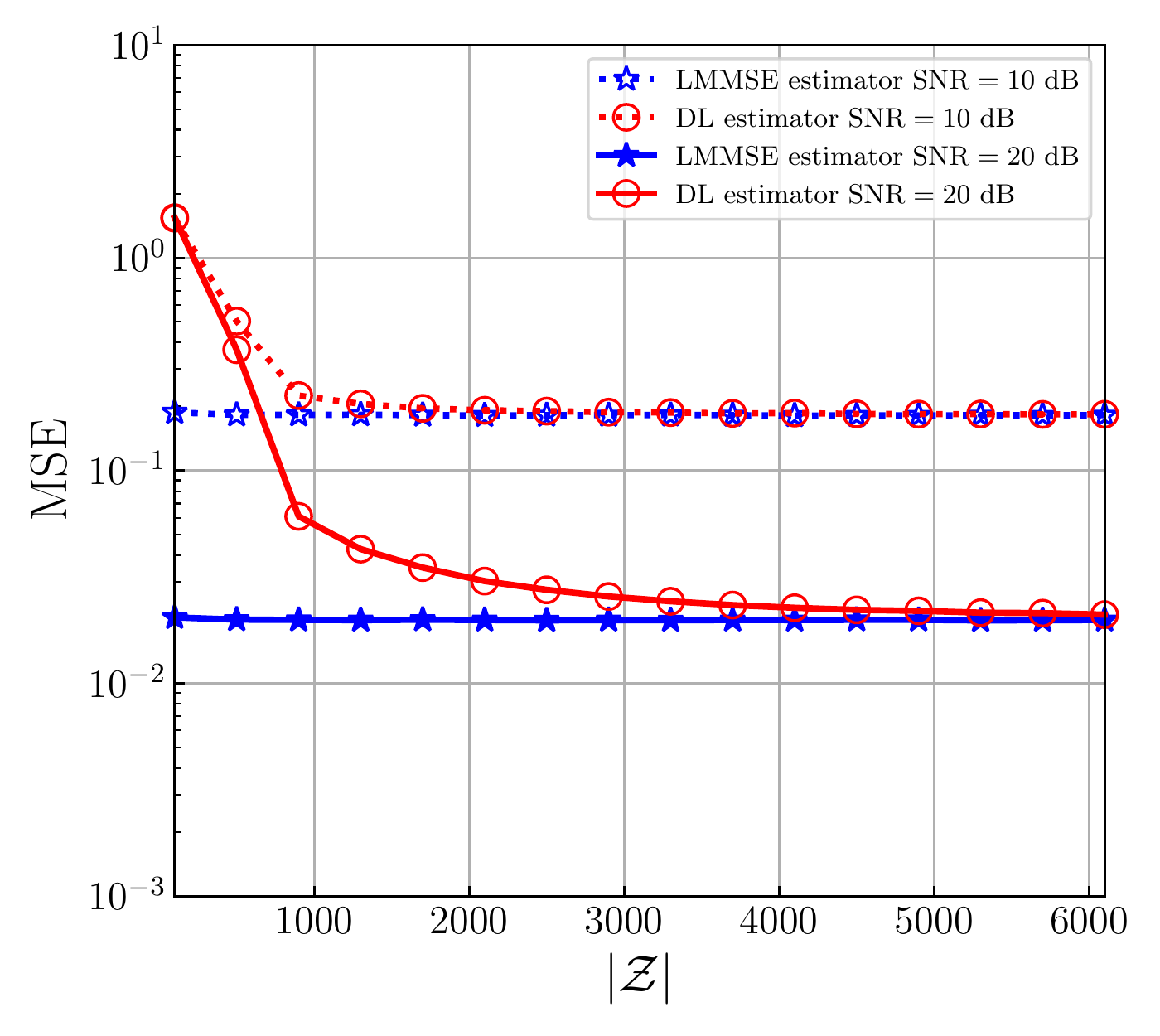}
		\caption{The MSE performance of the DL estimator versus $|\mathcal{Z}|$ under linear signal model.}
		\label{figure2-b}
	\end{figure}

	Fig.~\ref{figure2} compares the MSEs of the LS, LMMSE, and DL estimators versus SNR under linear signal model~\eqref{r_train}. The channel, $\mathbf{h}$, is assumed to be Gaussian with zero mean and element-wise unit variance. The sizes of training and test sets are $20,000$ and $5,000$, respectively. The underlying network of the DL estimator has $4$ layers and equal numbers of neurons at each hidden layer, i.e., equal widths. Denote $\tilde{d}$ as the width of hidden layer and $\tilde{d}$ is set to be $40$. From Fig.~\ref{figure2}, the MSEs of the LMMSE and the DL estimators are almost overlapped. Since the LMMSE estimator is equivalent to the MMSE estimator in this case, the DL estimator can well approximate $\mathbf{h}_{\mathrm{MMSE}}$, which confirms that $J(\mathbf{f}_{\boldsymbol{\theta}_{\mathcal{Z}}})\approx J_{\mathrm{LMMSE}}$ in the linear systems. Moreover, both the DL and LMMSE estimators outperform the LS estimator in Fig.~\ref{figure2} as noted by~\eqref{comparison}.
	
	Fig.~\ref{figure2-c} shows the MSEs of the DL estimator versus the width of ReLU DNN, $\tilde{d}$, under linear signal model~\eqref{r_train} with fixed SNRs and $d=2$. The MSEs of the LMMSE estimators derived under the same SNRs are used as the benchmark. When $\tilde{d}$ is small, the dimension of the parameter space $\Theta$ is very low and the approximation error becomes relatively high. As a result, the MSEs of the DL estimator are significantly larger than the MSEs of the LMMSE estimator. As $\tilde{d}$ increases, the parameter space $\Theta$ is enlarging and the approximation error decreases. All MSEs of the DL estimator at different SNRs approach to the MSEs of the LMMSE estimator. Such a result evidences the conclusions in Theorem~\ref{theorem1-1}.
	
	Fig.~\ref{figure2-b} shows the MSEs of the DL estimator versus the size of training samples, $|\mathcal{Z}|$, under linear signal model~\eqref{r_train} with fixed SNRs and $d=2$. As in~Fig.~\ref{figure2-b}, the LMMSE estimator is used as the benchmark. When $|\mathcal{Z}|$ is small, the generalization error is very high and the MSEs of the DL estimator are significantly larger than the MSEs of the LMMSE estimator. As $|\mathcal{Z}|$ increases, the generalization error decreases and all MSEs of the DL estimator at different SNRs  asymptotically approach to the MSEs of the LMMSE estimator. Such a result evidences the conclusions in Theorem~\ref{theorem1-1}.

	\subsection{Nonlinear Systems}	
	In this subsection, we evaluate the performance of the MMSE, LMMSE and DL estimators under a nonlinear signal model. Let $\mathbf{x}_{\mathrm{in}}=\mathbf{h}\tau+\mathbf{n}$ and the following nonlinear model
	\begin{align}\label{nonpa}
	x_{i}=x_{\mathrm{in},i}\bigg(1+\Big(\frac{x_{\mathrm{in},i}}{x_{\mathrm{sat}}}\Big)^{2\omega}\bigg)^{-\frac{1}{2\omega}}
	\end{align}
	is adopted,
	where $x_{i}$ and $x_{\mathrm{in},i}$ are the $i$-th elements of $\mathbf{x}$ and $\mathbf{x}_{\mathrm{in}}$, respectively, for $i\in\{1,\ldots,d\}$, $x_{\mathrm{sat}}$ is the saturation level, and $\omega$ is the smoothness factor. The other settings are the same as in Section~\ref{linear-simu}. The model in~\eqref{nonpa} is typically used by nonlinear signal detection caused by imperfection of PA and is commonly known as Rapp model~\cite{joung2014survey}. Here, we apply such a model to illustrate channel estimation for nonlinear systems. 
	
	\begin{figure}[!t]
		\centering
		\includegraphics[width=0.75\columnwidth]{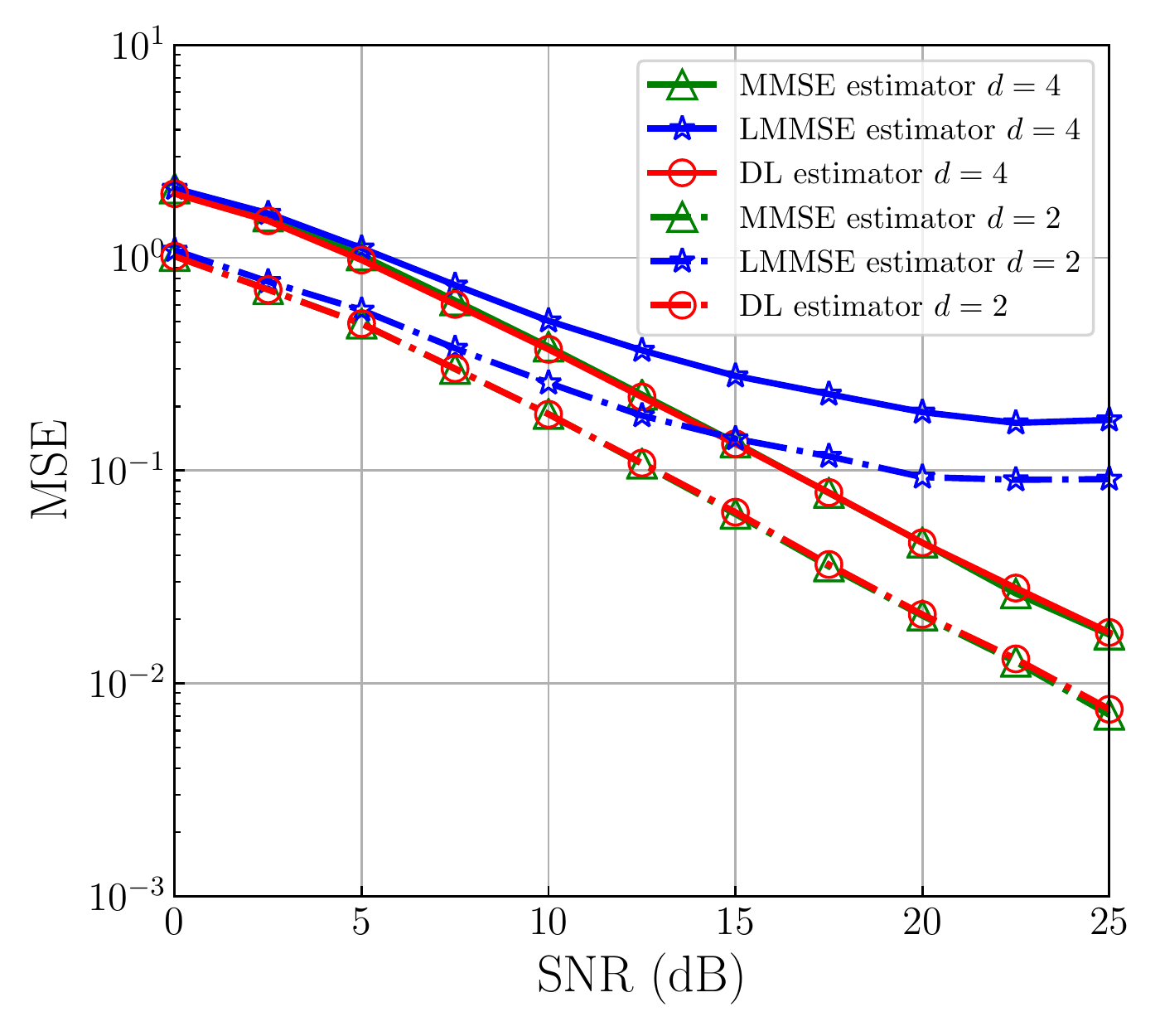}
		\caption{The MSE performance of the MMSE, LMMSE, and DL estimators versus SNR under nonlinear signal model.}
		\label{figure2-1}
	\end{figure}
	Fig.~\ref{figure2-1} shows the MSEs of the MMSE, LMMSE, and DL estimators versus SNR under nonlinear model in~\eqref{nonpa}, where the saturation level, $x_{\mathrm{sat}}$, is fixed as $1.5$ and the smoothness factor $\omega$ is set be $1$. Since no analytic form of $\mathbf{h}_{\mathrm{MMSE}}$ for nonlinear model~\eqref{nonpa} is available, we use Monte Carlo simulation to estimate $\mathbf{h}_{\mathrm{MMSE}}$ in Fig.~\ref{figure2-1} and the number of trials is set as $2\times 10^7$.  The performance of the MMSE, LMMSE, and DL estimators is close to each other at low SNRs as the noises dominate the overall MSEs. As the SNR increases, the approximation errors to the MMSE estimator will contribute a larger percentage of the MSEs. According to Theorem~\ref{theorem1}, the bias of the DL estimator is significantly lower than that of the LMMSE estimator for nonlinear systems. As a result, the performance of the DL estimator is very close to that of the MMSE estimator and becomes significantly better than that of the LMMSE estimator for high SNRs.

	\subsection{Robustness to Mismatched Information}
	We then compare the MSEs of channel estimation using the LMMSE and the DL estimators under inaccurate statistics of channel and mismatched training data in linear systems. Assume that $\mathbf{R}_{1}=\sigma_{1}^{2}\mathbf{I}_{d}$ and $\mathbf{R}_{\mathrm{er}}=\sigma_{\mathrm{er}}^{2}\mathbf{I}_{d}$, where $\sigma_{1}^{2}$ and $\sigma_{\mathrm{er}}^{2}$ are the element-wise variances. Moreover, we define the scaling coefficient $\eta$ as the ratio $\sigma_{1}^{2}/\sigma^{2}$ or $\sigma_{\mathrm{er}}^{2}/\sigma^{2}$. The other settings are the same as Section~\ref{linear-simu}. When $\eta>1$, i.e., Case I in Section~\ref{dlerror}, the performance of the DL estimator is only affected by the degree of  the mismatch for training data with real channel statistics. When $\eta<1$, i.e., Case II in Section~\ref{dlerror}, the DL estimator may malfunction and outputs random estimates due to the restricted effective input range.
	
\begin{figure}[t]
		\vskip 0.2in
		\centering
		\subfigure[Case I ($\eta=2>1$)]{
			\begin{minipage}[t]{0.46\linewidth}
				\centering
				\includegraphics[width=\columnwidth]{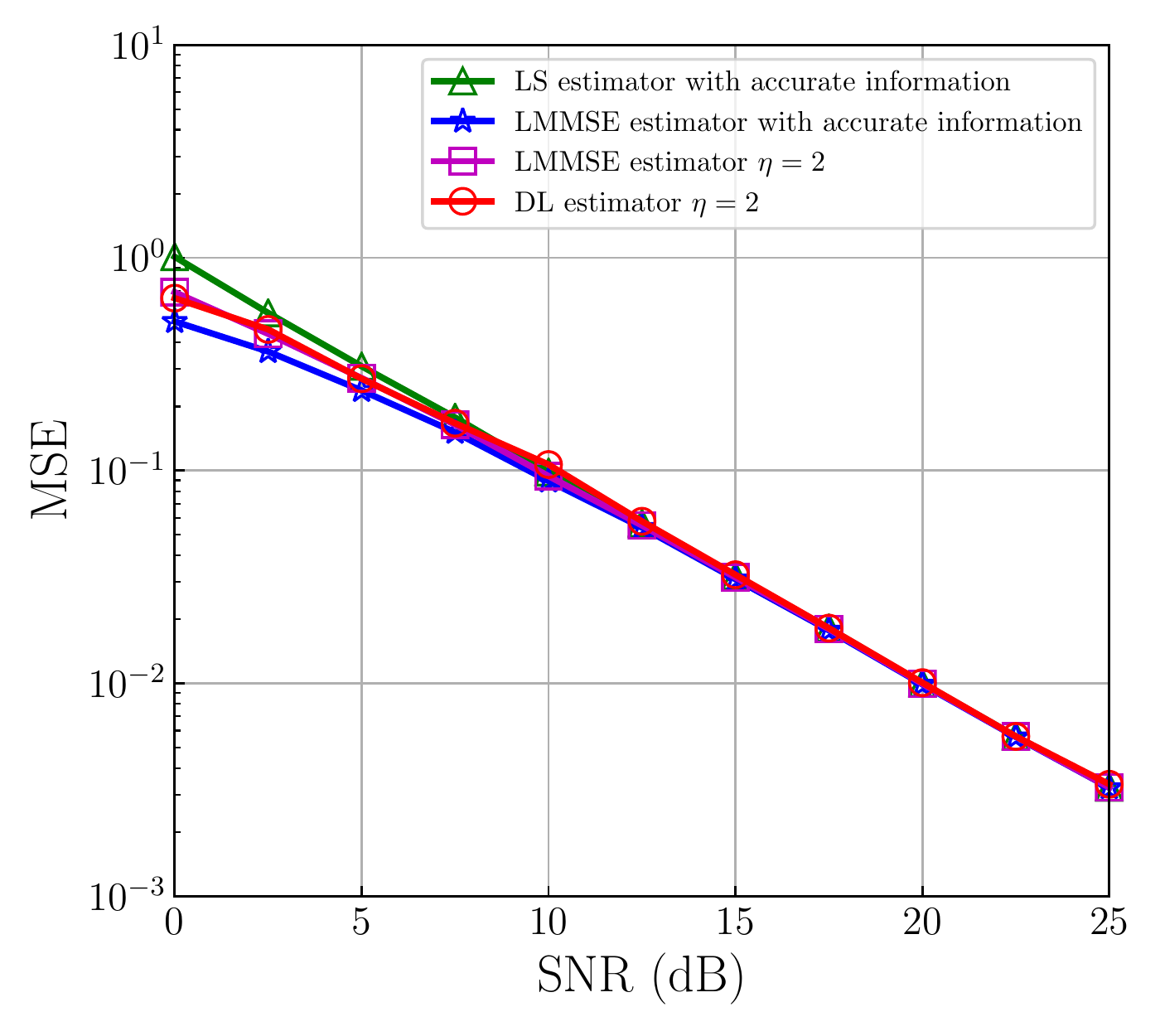}
			\end{minipage}
		}
		\subfigure[Case II ($\eta=0.2<1$)]{
			\begin{minipage}[t]{0.46\linewidth}
				\centering
				\includegraphics[width=\columnwidth]{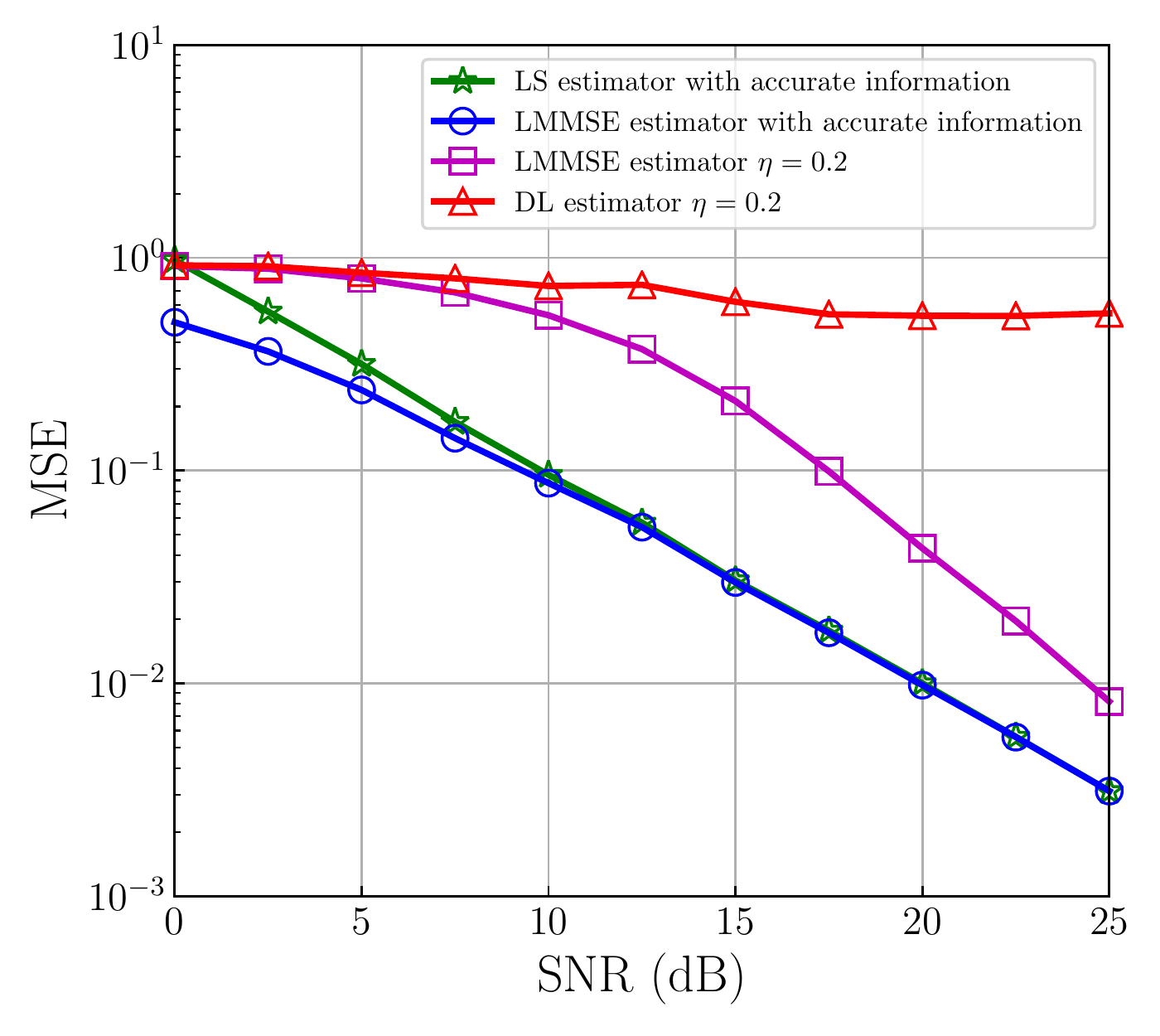}
			\end{minipage}
		}

		\caption{The MSE performance of the LMMSE and the DL estimators versus SNR under inaccurate statistics of channel and mismatched training data.}
		\label{figure3}
		\vskip -0.2in
	\end{figure}

	Fig.~\ref{figure3}(a) illustrates the MSEs of the LMMSE and the DL estimators versus SNR under linear signal model in~\eqref{r_train} with $d=1$ and $\eta=2$ that corresponds to Case I. The MSEs of the LMMSE estimator with accurate channel statistics and the LS estimator are served as the benchmarks. In Fig.~\ref{figure3}(a), both the LMMSE estimator with inaccurate channel statistics and the DL estimator with mismatched training data perform poorer than the LMMSE estimator with accurate channel statistics but still better than the LS estimator. Furthermore, under the same $\eta$, the MSEs of the LMMSE with inaccurate statistics and the DL estimator with mismatched training data are overlapped, which confirms~\eqref{lmmse1} and~\eqref{DL-ER}. Specifically, in high SNRs, the MSEs of these estimators are almost the same and the errors of channel statistics have little impact on the overall estimation performance.

		\begin{figure}[t]
		\vskip 0.2in
		\centering
		\subfigure[$\mathrm{SNR}=0\ \mathrm{dB}$]{
			\begin{minipage}[t]{0.46\linewidth}
				\centering
				\includegraphics[width=\columnwidth]{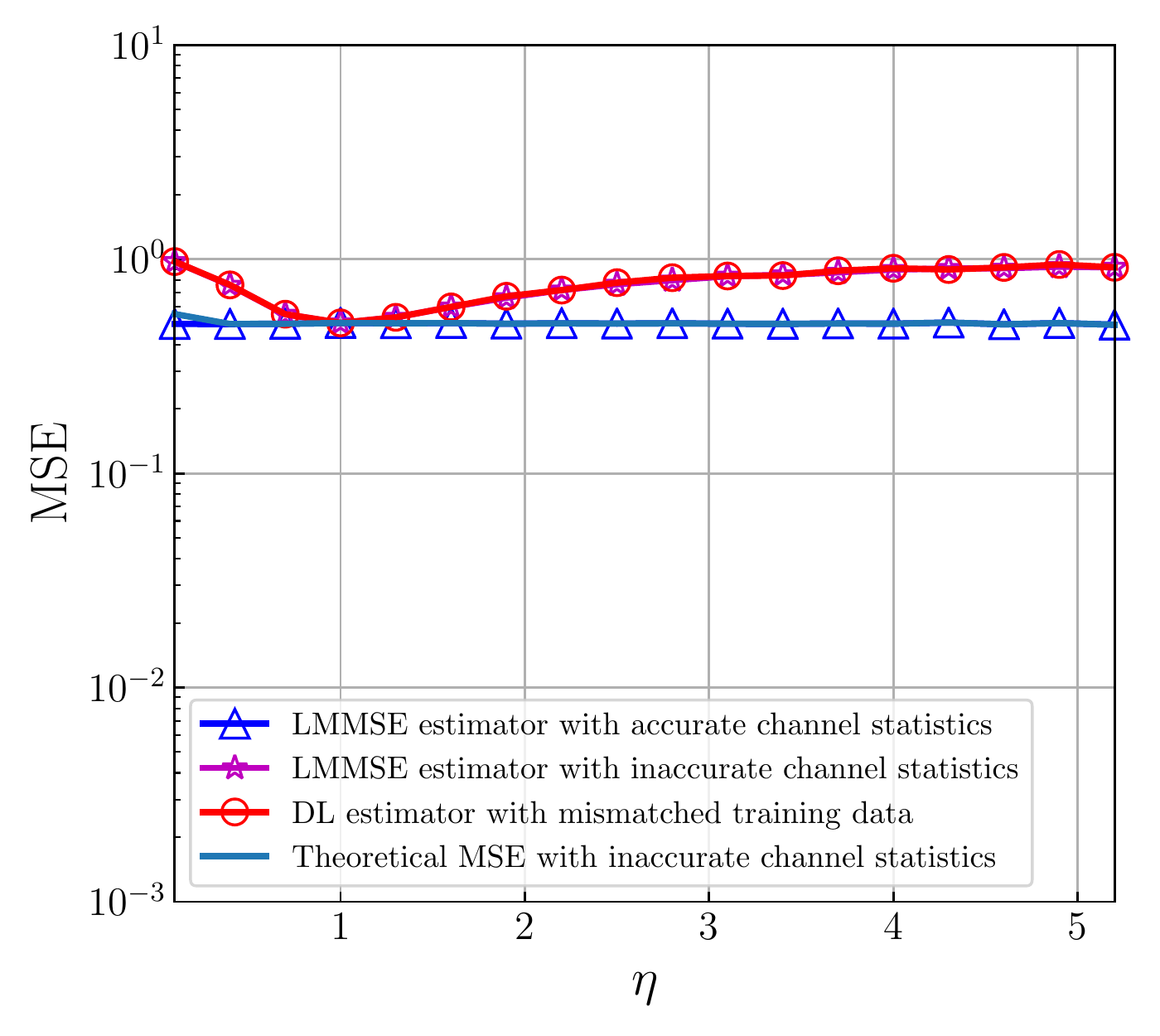}
			\end{minipage}
		}
		\subfigure[$\mathrm{SNR}=25\ \mathrm{dB}$]{
			\begin{minipage}[t]{0.46\linewidth}
				\centering
				\includegraphics[width=\columnwidth]{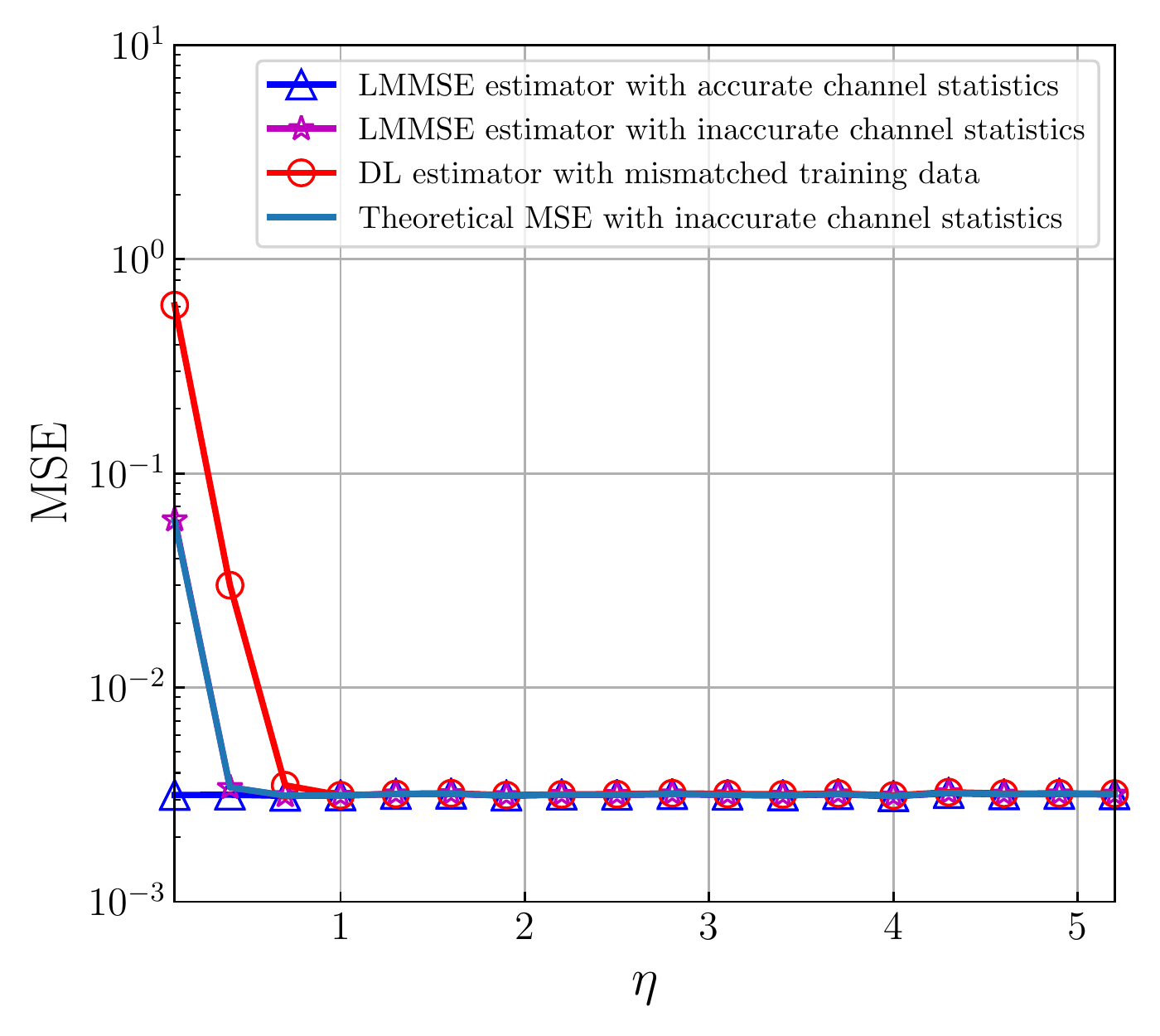}
			\end{minipage}
		}

		\caption{The MSE performance of the LMMSE and the DL estimators versus $\eta$ under inaccurate statistics of channel and mismatched training data.}
		\label{figure4}
		\vskip -0.2in
	\end{figure}
	
	Fig.~\ref{figure3}(b) illustrates the MSEs of the LMMSE and the DL estimators versus SNR under linear signal model in~\eqref{r_train} with $d=1$ and $\eta=0.2$ that corresponds to Case II. When $\eta< 1$, the MSE of the LMMSE estimator with inaccurate channel statistics is significantly larger than those of the LS and LMMSE estimators with accurate channel statistics. The performance of the LMMSE estimator with inaccurate channel statistics degrades more severely than Case I in Fig.~\ref{figure3}(a). The performance of the DL estimator is even worse since its MSE is totally random and uncorrelated to the SNR. Such phenomenon verifies the analysis in Section~\ref{dlerror} when the variance of training data is lower than that of true channel, i.e., $\eta <1$. Therefore, the mismatch of training data with the true environment is a serious problem if $\eta <1$ and should be carefully considered when applying DL methods in wireless communication systems.
	
	Fig.~\ref{figure4}(a) visualizes the MSEs of the LMMSE estimator with inaccurate channel statistics and the DL estimator with mismatched training data versus $\eta$ under linear signal model~\eqref{r_train} with $d=1$ and $\mathrm{SNR}=0\ \mathrm{dB}$. We adopt the MSE of the LMMSE estimator with inaccurate channel statistics in~\eqref{lmmse1} as the theoretical MSE. In Fig.~\ref{figure4}(a), the MSEs of the LMMSE estimator with inaccurate channel statistics and the DL estimator with mismatched training data are overlapped with the theoretical MSE and are slightly higher than the MSE of the LMMSE estimator with accurate channel statistics when $\eta> 1$.  Such a result verifies the correctness of~\eqref{lmmse1} and $\eqref{DL-ER}$, as $\eqref{DL-ER}$ is equivalent to~\eqref{lmmse1} under the same $\eta$. When $\eta=1$, there is no error. When $\eta< 1$, the MSEs of the LMMSE estimator with inaccurate channel statistics and the DL estimator with mismatched training data are larger than that of the LMMSE estimator with accurate channel statistics.  As illustrated in Fig~\ref{figure3}(b), the MSE of the LMMSE estimator with inaccurate channel statistics is comparable to that of the DL estimator with mismatched training data at low SNRs. Therefore, the gap between the MSEs of the LMMSE estimator with inaccurate channel statistics and the DL estimator with mismatched training data is not significant in Fig.~\ref{figure4}(a).
	
	Fig.~\ref{figure4}(b) shows the MSEs of the LMMSE estimator with inaccurate channel statistics and the DL estimator with mismatched training data versus $\eta$ when $d=1$ and $\mathrm{SNR}=25\ \mathrm{dB}$. When $\eta\geq 1$, the MSEs of the LMMSE estimator with inaccurate channel statistics and the DL with mismatched training data are almost equal to that of the LMMSE estimator. When $\eta<1$, the MSE of the DL estimator with mismatched training data is significantly larger than that of the LMMSE estimator with inaccurate channel statistics and the theoretical MSE. The reason for this phenomenon is that the estimate of the DL estimator with mismatched training data gets more random as $\eta$ decreases and is nearly uncorrelated to the SNR when $\eta<1$, as shown in Fig.~\ref{figure3}(b), while the MSE of the LMMSE with inaccurate channel statistics still decreases with the SNR. Therefore, the gap between the MSEs of the LMMSE estimator with inaccurate channel statistics and the DL estimator with mismatched training data becomes much more significant at high SNRs. Such a result verifies the analysis in Section~\ref{dlerror} again. Moreover, the MSE of the LMMSE estimator with inaccurate channel statistics matches the theoretical MSE and is slightly higher than the MSE of the LMMSE estimator with accurate channel statistics across the entire range of $\eta$, which shows the robustness of the LMMSE estimator to inaccurate channel statistics.

	\section{Conclusions}\label{conclusion-sec}
	
	In this paper, we have made the first attempt on interpreting DL based channel estimation under linear, nonlinear, and inaccurate channel statistics using a multiple antenna system as an example. We have explained that the DL estimator equipped with a ReLU DNN is mathematically equivalent to a piecewise linear function and can attain universal approximation to the MMSE estimator under suitably configured structure and large training samples. Extensive simulation results have confirmed the performance of the DL estimator and showed that the DL estimator is close to the LMMSE estimator under linear systems but significantly outperforms it when the signal model is nonlinear. However, the DL estimator is sensitive to the quality of training data and its performance would significantly degrade if the data in real environments distributes broader than the training data. The benefits of the DL estimator have to weigh against its costs when applied to the channel estimation in real wireless communication systems. We should strike a balance between DL based channel estimation and traditional channel estimation.

	
	
	\bibliography{inference}
	\bibliographystyle{IEEEtran}
	
\end{document}